\documentclass[]{jfm}

\usepackage{graphicx,color,rotating,psfrag,pstricks,pstricks-add}
\graphicspath{{plots/}} 
\usepackage{amssymb,amsmath,overpic,nomencl,array}
\usepackage{accents}
\usepackage{tabularx}
\newcolumntype{L}[1]{>{\raggedright\arraybackslash}p{#1}} 
\newcolumntype{C}[1]{>{\centering\arraybackslash}p{#1}} 
\newcolumntype{R}[1]{>{\raggedleft\arraybackslash}p{#1}} 

\newcommand{\vect}[1]{\boldsymbol{#1}}
\newcommand{\myred}[1]{\textcolor{black}{#1}}
\newcommand{\myredred}[1]{\textcolor{black}{#1}}
\usepackage{natbib}
\usepackage[colorinlistoftodos,disable]{todonotes} 
\graphicspath{{./plots/}}
\begin{document}
\title{Mean flow stability analysis of oscillating jet experiments}

\author[K.~Oberleithner {\itshape et al.}]{%
K\ls I\ls L\ls I\ls A\ls N \ns O\ls B\ls E\ls R\ls L\ls E\ls I\ls T\ls H\ls N\ls E\ls R$^1$, \textcolor{black}{L\ls O\ls T\ls H\ls A\ls R \ns R\ls U\ls K\ls E\ls S $^2$}  \and J\ls U\ls L\ls I\ls O\ns S\ls O\ls R\ls I\ls A$^{1,3}$}
\affiliation{$^1$Laboratory for Turbulence Research in Aerospace \&
Combustion\\ Monash University\\ Melbourne, VIC 3800, Australia\\[\affilskip]
\textcolor{black}{$^2$Institut f\"ur Str\"omungsmechanik und Technische Akustik, HFI\\ Technische Universit\"at Berlin\\ 10623 Berlin, Germany}\\[\affilskip]
$^3$Department of Aeronautical Engineering\\
King Abdulaziz University\\
Jeddah, Kingdom of Saudi Arabia
}

\maketitle
\begin{abstract}
\myred{
Linear stability analysis is applied to the mean flow of an oscillating round jet with the aim to investigate the robustness and accuracy of mean flow stability wave models.
The jet's axisymmetric mode is excited at the nozzle lip through a sinusoidal modulation of the flow rate at amplitudes ranging from $0.1~$\% to $100~$\%.
The instantaneous flow field is measured via particle image velocimetry and decomposed into a mean and periodic part utilizing proper orthogonal decomposition. 
Local linear stability analysis is applied to the measured mean flow adopting a weakly nonparallel flow approach. 
The resulting global perturbation field is carefully compared to the measurements in terms of spatial growth rate, phase velocity, and phase and amplitude distribution. 
It is shown that the stability wave model accurately predicts the excited flow oscillations during their entire growth phase and during a large part of their decay phase. 
The stability wave model applies over a wide range of forcing amplitudes, showing no pronounced sensitivity to the strength of nonlinear saturation. 
The upstream displacement of the neutral point and the successive reduction of gain with increasing forcing amplitude is very well captured by the stability wave model. 
At very strong forcing ($>40\%$), the flow becomes essentially stable to the axisymmetric mode. For these extreme cases, the prediction deteriorates from the measurements due to an interaction of the forced wave with the geometric confinement of the nozzle. 
Moreover, the model fails far downstream in a region where energy is transferred from the oscillation back to the mean flow. 
This study supports previously conducted mean flow stability analysis of self-excited flow oscillations in the cylinder wake and in the vortex breakdown bubble and extends the methodology to externally forced convectively unstable flows. The high accuracy of mean flow stability wave models as demonstrated here is of great importance for the analysis of coherent structures in turbulent shear flows.}
 
\end{abstract}
\section{Introduction}
\label{sec:intro}
Instabilities inherent to turbulent shear flows cause small perturbations to grow significantly in space and time. 
This leads to the formation of large-scale coherent flow structures that play an important role for cross-flow momentum transfer, noise generation, and mixing. 
The active control of shear flows is most efficient if these inherent instabilities are exploited \citep{Greenblatt2000}. 
\par
In the last decades, significant effort has been made to derive analytic models for coherent structures and their impact on the mean and turbulent flow characteristics. The underlying methodology is typically based on a triple decomposition of the time dependent flow into a time-mean part, a \myred{periodic (coherent)} part, and a randomly fluctuating (turbulent) part.
\myred{The mean is obtained from time-averaging, while the periodic part is obtained from a phase average,``i.e. the average over a large ensemble of points having the same phase with respect to a reference oscillator''\citep{Reynolds1972}. The coherent part represents fluctuations at large time and length scales in contrast to the small-scale turbulent fluctuations. This separation of scales allows for the treatment of the wave-like coherent structures independently from the random turbulent fluctuations.}  
\par
\myred{Since the groundbreaking experiments of \cite{Crow1971} and  \cite{Brown1974}, many researches have shown that the large-scale coherent structures in open shear flows are qualitatively similar to instability waves. As stated by \cite{Gaster1985}, ``This similarity between the patterns in laminar and turbulent states is not very surprising in view of the fact that the basic long-wave vorticity-transport instability mechanism is mainly controlled by the mean-velocity profiles of the flow, and these are not too different in the two situations.''.}
\par
\myred{This phenomenological reasoning leads to instability wave models, where the coherent structures are derived from a linear stability analysis (LSA) of the mean turbulent flow. 
Since these models are based on the (nonlinearly modified) mean flow they intrinsically account for the mean--coherent and mean--turbulent interactions.} The turbulent-coherent interactions are not represented by the mean flow and are either neglected \citep{Gaster1985,Cohen1987a,Gudmundsson2011,Oberleithner2011a} or lumped into an eddy viscosity model \citep{Marasli1991,Reau2002b,Oberleithner2013c}. 
\cite{Reau2002b} and \cite{Lifshitz2008} developed a mathematical model that incorporates all the three interactions for the case of the forced turbulent mixing layer with the goal of turbulent-coherent closure. Their model captures the interaction between the mean flow and the coherent structures qualitatively, but the actual growth rates deviate from the measurements, particularly for higher forcing amplitudes. \cite{Lifshitz2008} suggest that the erroneous prediction is attributed to an inaccurate model of the turbulent-coherent interactions. 
This conclusion is stereotypical for the LSA of turbulent flows, where the inaccurate prediction of the large-scale structures is typically related to a number of causes that are difficult to separate. 
Reasons for an inaccurate prediction can be an insufficient turbulent-coherent interaction model, an inaccurate mean flow representation (e.g. accidental modification through finite amplitude forcing), strong nonparallelism of the flow (for local stability analysis), or nonlinear mode-mode interaction. 
Due to the multitude of error sources, the stability wave models in turbulent flows are still considered as inaccurate and are far from being generally accepted. 
\par
Mean flow stability wave models have significant importance in the analysis of experimental work.
While LSA based on the unperturbed state is useful to predict the onset of instability, the mean flow LSA shows a great potential for the analysis of the nonlinearly saturated state that manifests in a (perturbed) experimental environment. 
The most prominent example of successful mean flow LSA is the prediction of the finite vortex shedding in the cylinder wake \citep{Pier2002,Barkley2006}. For a wide range of post-critical Reynolds numbers, the frequency at the limit-cycle is precisely predicted by the linear global mode of the mean flow, while the prediction based on the steady state solution deteriorates with increasing distance from the critical point. The results are perfectly in line with the mean field model proposed by \cite{Noack2003}. Yet, the validity of the mean flow LSA is not rigorous and depends on the interactions between the mean flow, the fundamental, and its harmonics. As demonstrated by \cite{Sipp2007a}, the mean flow LSA is inaccurate in the case of a cavity flow, due to a strong resonance of the fundamental wave with its first harmonic.  In this work we will clarify whether this mechanism is relevant for the axisymmetric jet. 
\par
In this investigation, \myred{we consider an} essentially laminar flow that is subjected to external forcing causing a finite amplitude wavetrain to grow and decay in the streamwise direction due to convective hydrodynamic instability. The primary interactions that take place are between the mean flow and the oscillation and between the oscillation and its harmonics. 
LSA is applied to the resulting time-averaged flow obtained from measurements with the aim to predict the oscillating flow field. 
By changing the amplitude of the forcing, we may control the intensity of the nonlinear interactions involved in the saturation process. 
With this approach, we can analyze to what extent the stability wave models are affected by these interactions, without the ambiguity of a turbulence model affecting the conclusions.
\par
Although, \cite{Pier2002} and \cite{Barkley2006} \myred{have already demonstrated} an excellent prediction of the vortex shedding frequency based on the mean flow LSA, these authors did not compare the mode shapes with the actual oscillatory field, and the ability of the mean flow LSA in this regard remains unknown. 
In the present study, we undertake a comprehensive comparison of the oscillating flow field with the LSA predictions, considering growth rates, phase velocities, and phase and amplitude distributions.
\textcolor{black}{Moreover, we do not restrict ourselves to the class of self-excited flows that are driven by a global instability, but we consider the, possibly, more general configuration of a globally stable but convectively unstable flow.} 
For the forced jet flow considered here, perturbations are initiated at the nozzle at precisely controlled conditions, and they are convected downstream while growing and decaying at a rate that is determined by the underlying base flow stability. Hence, the amplitude of the instability wavetrain is rather a function of space than time, and so is the mean-coherent interaction. The same concept applies to the flow bifurcating to a global mode, however in that scenario, the location of the (internal) forcing at the initiated onset of instability and at the limit-cycle is less precisely known. In fact, the local LSA adopted in this work can be rigorously applied to both flow classes, with the difference that for the globally unstable flows, the forcing frequency and location must be derived through a spatio-temporal type analysis of the flow field, while for the convectively unstable flows, it is readily defined by the type of forcing applied (see e.g. \cite{Huerre1990,Juniper2011a}). 
\par
The globally stable but convectively unstable flow considered  here is suitable for a local stability analysis. 
We adopt the weakly nonparallel flow analysis developed by \cite{Crighton1976}. Within the framework of multiple-scale analysis, the local amplitude function is assumed to vary slowly in the streamwise direction, allowing for an unambiguous reconstruction of the overall flow response to a localized perturbation. The same approach has been used for the inviscid analysis of turbulent jets \citep{Strange1983} and mixing layers \citep{Gaster1985,Lifshitz2008} and the viscous analysis of laminar swirling jets \citep{Cooper2002}. 
\par
The remainder of the manuscript is organized in the following way. 
The experimental methods are summarized in \S \ref{sec:exp_method}, including a description of the laminar jet-rig, the particle image velocimetry (PIV) experiments, and the adopted phase-reconstruction scheme.
The adopted mean flow stability wave model is described in \S \ref{sec:theory}, providing a short derivation of the mean flow stability equations and the solution ansatz for the weakly nonparallel jet. 
\S \ref{sec:AmpVariation} provides an overall description of the effect of forcing on the streamwise growing instability waves, while a detailed comparison of the experimental data with the stability wave model is given in \S \ref{sec:detailed_comparison}. The result sections conclude with an energy flux analysis given in \S \ref{sec:interaction}, and the main observations are summarized in \S \ref{sec:conclusions}.
\section{Experimental method}
\label{sec:exp_method}
\subsection{Experimental setup and procedure}
\begin{figure}
\centering
\input{plots/VortexRingRig2.tex}
\fbox{
\includegraphics[width=.9\textwidth]{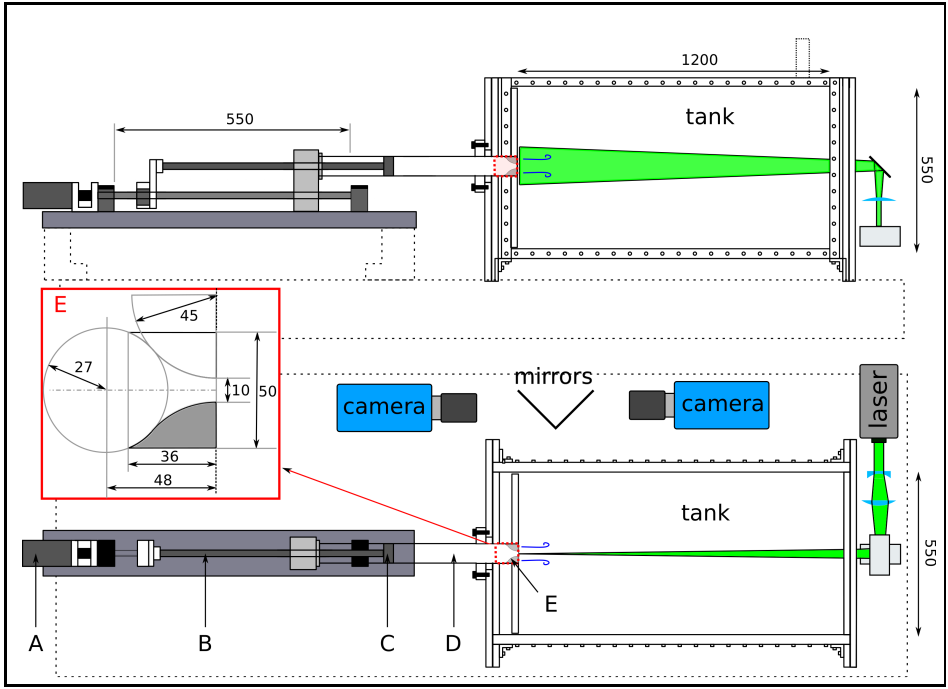}}
\caption{Side and top view of the jet facility (all dimensions expressed in mm); The capital letters refer to the stepper motor (A), the lead screw (B), the piston (C), the tube (D), and the contracting nozzle (E).}
\label{fig:Setup} 
\end{figure}

Figure \ref{fig:Setup} shows a schematic diagram of the experimental apparatus used in this study, with water as the working fluid. 
The same facility has been used in previous studies examining zero-net-mass-flux jets \citep{Cater2002} and continuous low Reynolds number round jets \citep{ONeill2004}.
The fluid motion is generated by a piston traversing through a $700~$mm long perspex tube with an inner diameter of $50~$mm.
The fluid is guided through a smooth contracting nozzle mounted to the end of the tube before it is discharged into a large water tank. 
The contraction is made of stainless steel and terminates at a diameter of $D=10~$mm.  To achieve 'clean' upstream geometric boundary conditions, a fake wall is inserted into the tank that is flush with the nozzle lip. To avoid surface waves, the tank is closed from the top using a perspex roof with a riser tube
with an inner diameter of 56.5 mm located at the far end of the tank, and the facility is filled up to the roof.
A lead screw is used to transfer the rotational motion of the stepper motor to the transversal motion of the piston, yielding a resolution of $0.2~\mu$m per step. The motor motion is closed-loop controlled by feeding an encoder signal back to the motor driver. 
The motor driver allows for running precompiled motion profiles at an update rate of $2~$ms.
\par
Experiments were conducted at a facility Reynolds number of $Re_D=U_{\mathrm{jet}}D/\nu=770$, with $D$ being the nozzle diameter,  $U_{\mathrm{jet}}$ the plug flow velocity at the nozzle, and $\nu$ the kinematic viscosity of water. 
The jet is forced axisymmetrically by imposing a sinusoidal motion onto the mean motion of the piston\myred{, yielding a piston velocity,
\begin{equation}\label{eq:Upiston}
 u_{\mathrm{piston}} = \overline u_{\mathrm{piston}}\left(1+A\cos(2\pi ft)\right),
\end{equation}
with the excitation amplitude defined as
\begin{equation}\label{eq:Apiston}
 A = \hat u_{\mathrm{piston}}/\overline{u}_{\mathrm{piston}},
\end{equation}
where $\hat u_{\mathrm{piston}}$ refers to the amplitude of the oscillating motion of the piston, $\overline{u}_{\mathrm{piston}}$ to the time-averaged velocity of the piston, and $f$ to the excitation frequency.} The amplitude of the piston motion is considered equivalent to the amplitude of the axial velocity fluctuations at the nozzle.
The phase-locked velocity fluctuations measured at the nozzle confirm this assumption. 
For this study, data was recorded for forcing amplitudes ranging from $A=0~$\% to $A=100~$\%. 
\textcolor{black}{The excitation frequency was $f=2~$Hz yielding a Strouhal number of $fD/U_{\mathrm{jet}}=0.26$. The frequency was selected from a stability analysis of the natural flow, with the aim of exciting an instability that reaches neutral stability in the center of the measurment domain to capture its entire growth and decay phase. The frequency does not correspond to the mode with the largest overall amplification. It is worth noting that the axisymmetric jet is generally unstable to a wide range of modes with the axisymmetric mode being dominant in the potential core while the single-helical mode takes over further downstream \citep{Cohen1987a,Oberleithner2013c}. The present focus on the axisymmetric mode is arbitrarily motivated by the facility design, but we expect a similar general scenario when forcing the jet at a helical mode.}
\par
Before each experiment, the water in the tank was stirred using an aquarium pump, and the piston was moved to its most upstream position. After five more minutes, the flow in the tank was considered as \myred{stagnant} and the forward motion of the piston was initiated. A statistically stationary flow was established after approximately $30$ seconds and flow measurements were conducted after one minute. \myred{This provides more than two minutes for data acquisition before the end of the piston motion is reached}. 
\subsection{Flow measurements}
Planar particle image velocimetry (PIV) was used for the flow velocity measurments.
This image-based method enables the instantaneous measurement of the two components of the fluid velocity in the plane of the light sheet generated by a pulsed high-energy light source. 
Figure \ref{fig:Setup} shows the arrangement used in this experiment.
\par
Table \ref{tab:PIV} lists the important parameters of the PIV acquisition and analysis. 
The instantaneous PIV velocity fields were acquired at low frame rates and can therefore be considered to be uncorrelated. The single-exposed image pairs were analyzed using the multigrid cross-correlation digital PIV algorithm described in \cite{Soria1999}, which has its origin in an iterative and adaptive cross-correlation algorithm \citep{Soria1994c,Soria1996c,Soria1996b}. Details of the performance, precision, and experimental uncertainty of the algorithm with applications to the analysis of single exposed PIV and holographic PIV images have been reported by \cite{Soria1998b} and \cite{vonEllenrieder2001}, respectively. The current implementation included window deformation \citep{Huang1993} and $3\times3$ points least-squares Gauss peak fitting \citep{Soria1996b} to determine the velocity to sub-pixel accuracy in addition to B-spline reconstruction. This algorithm can resolve displacements as small as $0.1 \mbox{px} \pm 0.06 \mbox{px}$, (at the $95~$\% confidence level) \citep{Soria1996b}, which corresponds to an uncertainty of the instantaneous velocity of approximately $0.5~$\%. The velocity vectors obtained from PIV represent average values within the interrogation window and the laser sheet thickness, which causes an underestimation of the actual velocity in the presence of velocity gradients.  This is a potential error source for the stability analysis that is based on the measured mean flow, as the stability wave growth rates are related to the shear layer thickness. However, a preliminary study of an analytic velocity model showed that, for the present study, these smoothing effects have only very little impact on the growth rates derived from the stability analysis (less than $0.01$ variation in $\alpha_iD$).
\begin{table}
	\centering
	\begin{tabular}{lll}
		& dimensional & non-dimensional\\
\hline
final interrogation window & $24~$px$\times32~$px & $0.037D\times0.05D$\\
laser sheet thickness   & $<1~$mm & $<0.1D$\\
field of view   & $35~$mm$\times100~$mm & $3.5D\times10D$\\
laser pulse delay & $3$ms & $0.04D/U_{\mathrm{jet}}$\\
acquisition frequency & $1.15~$Hz &  $0.19U_{\mathrm{jet}}/D$\\
\hline
	\end{tabular}
	\caption{Parameters of PIV measurements}
	\label{tab:PIV}
\end{table}
\subsection{Decomposition of the velocity field}\label{sec:Decomp}
\myred{The flow field is expressed in cylindrical coordinates, with $x$ having its origin at the nozzle lip and being aligned with the axis of rotation, with $r=0$ representing the jet centerline, and with $\theta$ pointing in positive direction according to the right-hand rule. 
The velocity components in the direction of the coordinates $\vect x = (x,r,\theta)^T$ are $ \vect u = (u,v,w)^T$.}
\par
The flow field $\vect u(\vect x,t)$, which features a dominant oscillatory pattern, can be decomposed into a time-averaged (mean) part, a \myred{periodic (coherent)} part, and a remainder of these two, reading 
\begin{equation}\label{eq:triple_decomp}
\vect u(\vect x,t) = \overline{\vect u}(\vect x)+\widetilde{\vect u}(\vect x,t)+\vect u''(\vect x,t).
\end{equation}
\myred{The time average is defined as 
\begin{equation}\label{eq:time_ave}
\overline{\vect u}(\vect x)=\lim\limits_{T \rightarrow \infty}{\frac{1}{T}}\int_0^T\vect u(\vect x,t)\mathrm{d}t.
\end{equation}
The wave (coherent) component \myred{$\widetilde{\vect u}$} is obtained from subtracting the mean flow from the phase-averaged flow field, reading  
\begin{equation}\label{eq:coherent}
\widetilde{\vect u}(\vect x,t)=\langle{\vect u}(\vect x,t)\rangle-\overline{\vect u}(\vect x).
\end{equation}
where the phase-average is defined as 
\begin{equation}\label{eq:phase_ave}
\langle{\vect u}(\vect x,t)\rangle=\lim\limits_{N \rightarrow \infty}{\frac{1}{N}}\sum\limits_{n=0	}^N\vect u(\vect x,t+n\tau),
\end{equation}
with $\tau$ representing the period of the wave.}
\par
To study the nonlinearities involved in the saturation of the \myred{excited} waves, it is convenient to decompose the coherent velocity field into the part oscillating at the fundamental frequency $f$ and its harmonics, yielding
$$
\widetilde{\vect u} = \widetilde{\vect u}_f+\widetilde{\vect u}_{2f}+\widetilde{\vect u}_{3f}+...
$$
\myred{
with the corresponding complex shape function 
\begin{equation}\label{eq:shape_function}
\hat{\vect u}_{(nf)}(\vect x) = \frac{1}{2\pi T}\int_0^T\widetilde{u}~\mathrm{e}^{-i2\pi nft}\mathrm{d}t, \ \ \ \mathrm{with} \ \ \ n=1,2,3...
\end{equation}
}
\todo{still not happy with this equation}
Fourier analysis is the method of choice to extract these quantities if time-resolved data is available. If the flow field is given as an ensemble of uncorrelated snapshots taken at arbitrary time increments, as it is the case for the present study, Proper Orthogonal Decomposition (POD) allows for a {\itshape a posteriori} reconstruction of the phase-averaged flow.  
This method is only applicable if a pair of POD modes can be identified that span the subspace of the oscillatory motion. 
\myred{The same method was previously applied to reconstruct the dominant coherent structures in supersonic jets \citep{Mitchell2014} and in jets with swirl \citep{Oberleithner2011a,Stoehr2011a,Oberleithner2012b}.}
In the present study, the fundamental and higher order harmonics clearly pair in POD space with stepwise decreasing energy content. In appendix \ref{sec:app2}, the phase-reconstruction scheme is demonstrated on the jet forced at $A=5~$\%. 
\par
\begin{figure}
\centering
\input{plots/POD_Vorticity_Contours_A5d0.tex}
\includegraphics[width=1\textwidth]{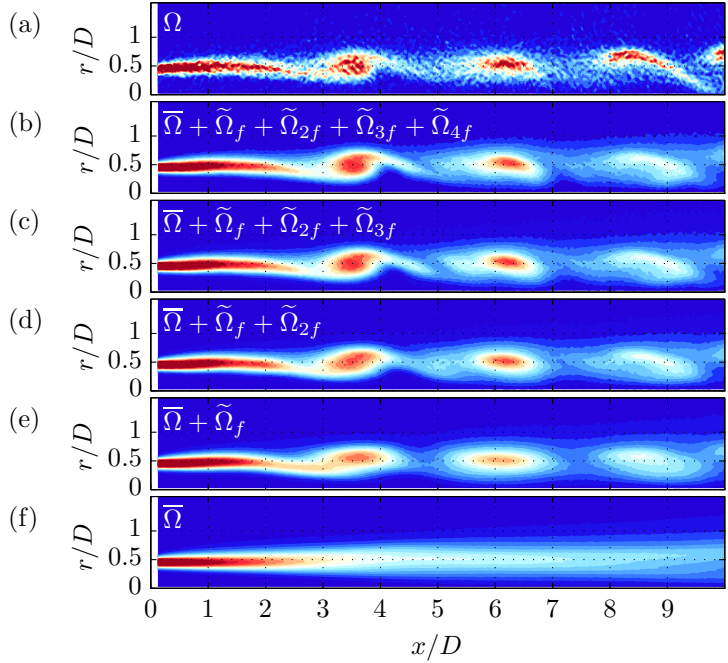}
\caption{Azimuthal vorticity field of the jet forced at $A=5~$\%: Instantaneous \myred{flow field} (a), phase-averaged flow field reconstructed from a decreasing number of harmonics (b-e), mean flow field (f). \myred{Instantaneous and phase-averaged flow fields are shown at the same phase reference.}}
\label{fig:POD_Vorticity_Contours_A5d0} 
\end{figure}
For demonstration purpose, figure \ref{fig:POD_Vorticity_Contours_A5d0} shows the reconstructed vorticity field for the jet forced at $A=5~$\%.
The contours \myred{shown in figure \ref{fig:POD_Vorticity_Contours_A5d0}a} represent the azimuthal vorticity, $\Omega=\partial v/\partial x-\partial u/\partial r$, of the instantaneous flow field obtained from a single PIV snapshot. 
The streamwise growth of the excited instability is clearly visible from the instantaneous velocity field. 
The roll-up of the shear layer into a single vortical structure occurs at around $x/D=4$ resulting in a strong agglomeration of vorticity that subsequently decays with downstream distance. 
The thin vortex sheet in between the regions of strong vorticity, sometimes referred to as the 'braid', is stretched during the roll-up process. 
Very similar structures are observed in the forced plane mixing layer reported by \cite{Weisbrot1988}. 
The contours shown in the adjacent frames show the same flow field reconstructed from a decreasing number of harmonics. 
It is apparent that the main oscillatory pattern is resolved by the fundamental frequency oscillations (first two POD modes), while the finer details during the roll-up of the shear layer are resolved by the higher order harmonics.  It is worth noting that the braid region is not resolved if the higher order harmonics are neglected, although they only contribute to a fraction of the total kinetic energy. 
\subsection{Parametrization of the mean flow}
The mean flow field is used as an input for the stability wave model. To avoid numerical difficulties stemming from the use of non-smooth data, we approximate the velocity profile of the axial component by the analytic expression
\begin{equation}\label{eq:Velfit}
\overline{u}(\textcolor{black}{x},\eta)/U_{\mathrm{cl}}= 0.5\left(1-\mathrm{tanh}~\eta\left[1+\mathrm{sech}^2\eta(C_1\mathrm{\tanh}~\eta+C_2)\right]\right),
\end{equation}
\begin{equation}
\ \ \ \ \ \ \text{with} \ \ \  \eta\textcolor{black}{(x)} = 0.5C_3(r-R_{.5})/\delta,
\end{equation}
where $U_{\mathrm cl}$ represents the axial velocity  on the jet centerline, $R_{.5}$ the half velocity radius of the jet, and
$\delta$ the shear layer momentum thickness defined as 
\begin{equation}\label{eq:momthick}
 \delta\textcolor{black}{(x)} = \int_0^{\infty}\frac{\overline{u}}{U_{\mathrm cl}}\left(1-\frac{\overline{u}}{U_{\mathrm cl}}\right)\mathrm{d}r.
\end{equation} 
This family of profiles was proposed by \cite{Cohen1987a}, where the constants $C_1$ and $C_2$  describe  the  respective symmetric  and 
antisymmetric corrections to the classical hyperbolic tangent profile, and $C_3$ represents the divergence of the center of the mixing layer from the centerline of the jet.
The constants are obtained \myred{for each streamwise station} from a least-square fit to the experimental data. To achieve smooth data, ninth-order polynomial functions are then fitted to the streamwise distribution of the constants. A smooth mean radial velocity component is derived from the mean axial velocity component by integrating the continuity equations.

\section{Stability wave model}
\label{sec:theory}
This section outlines the theoretical framework employed for the stability wave model. 
We first derive the equations that apply to infinitesimal perturbations traveling on a nonlinearly corrected  mean flow and then formulate the solution ansatz for a weakly-nonparallel axisymmetric jet flow. Note that two different order parameters are involved, one quantifying the perturbation amplitude and the other quantifying the nonparallelism of the mean flow. 
\subsection{\myred{Perturbation equations linearized around the mean flow}}
We start with the governing equations for an incompressible Newtonian fluid. These are the Navier--Stokes and continuity equations that are written in vector form as 
\begin{subequations}\label{eq:gov_equations}
\begin{align}
\frac{\partial{\vect u}}{\partial{t}} + \vect u\cdot\nabla\vect u &= -\nabla p +\frac{1}{\mathrm{Re}}\nabla^2\vect u \\
 \nabla\cdot\vect u &= 0 .
\end{align}
\end{subequations}
\par 
To analyze the \myred{linear} stability of the steady state, the velocity and pressure field are expressed as a sum of a steady and an infinitesimal perturbation field, yielding
\begin{equation}\label{eq:baseflow_decompsotion}
 \vect u = \vect u_b + \vect u' \ \ \text{and} \ \  p=p_b+p',
\end{equation}
where the base flow $\vect u_b$ is by definition a steady solution of \eqref{eq:gov_equations}.
The perturbation equations are obtained by substituting \eqref{eq:baseflow_decompsotion} into \eqref{eq:gov_equations}, subtracting the base flow equations, and neglecting the nonlinear term $\vect u'\cdot\nabla \vect u'$, yielding
\begin{subequations}\label{eq:gov_equations_perturbation1}
\begin{align}
\frac{\partial{\vect u'}}{\partial{t}} + \vect u'\cdot\nabla\vect u_b + \vect u_b\cdot\nabla\vect u'&= -\nabla p' +\frac{1}{\mathrm{Re}}\nabla^2\vect u' \\
 \nabla\cdot \vect u' &= 0 .
\end{align}
\end{subequations}
This set of equations describes the initial departure of a linearly unstable flow from its steady state before nonlinear saturation processes become active.
To analyze the saturated \myred{oscillating} state, the velocity and pressure fields are decomposed into a time-averaged and a (finite amplitude) \myred{periodic} part, yielding \begin{equation}\label{eq:double_decomposition}
 \vect u = \overline{\vect u} + \widetilde{\vect u} \ \ \text{and} \ \  p=\overline{p}+\widetilde{p} .
\end{equation}
Substituting this ansatz into the governing equations \eqref{eq:gov_equations} and taking the time-average leads to the governing equations for the mean flow, yielding
\begin{subequations}\label{eq:gov_equations_meanflow}
\begin{align}
\overline{\vect u}\cdot\nabla \overline{\vect u} &= -\nabla \overline {p} +\frac{1}{\mathrm{Re}}\nabla^2\overline{\vect u} + \vect F\\
 \nabla\cdot\overline{\vect u} &= 0,
\end{align}
\end{subequations}
where the forcing term $\vect F(\vect x)=-\overline{\widetilde{\vect u}\cdot \nabla \widetilde{\vect u}}$ represents the Reynolds stress induced by the flow field oscillations.
The mean flow is not a steady solution of \eqref{eq:gov_equations}, but it is a steady solution of the forced Navier--Stokes equations
\begin{subequations}\label{eq:gov_equations_forced}
\begin{align}
\frac{\partial{\vect u}}{\partial{t}} + \vect u\cdot\nabla\vect u &= -\nabla p +\frac{1}{\mathrm{Re}}\nabla^2\vect u +\vect F\textcolor{black}{^*} \\
 \nabla\cdot\vect u &= 0.
\end{align}
\end{subequations}
To analyze the linear stability of the mean flow, the velocity and pressure fields are expressed as a sum of the mean flow solution and an infinitesimal perturbation, yielding 
\begin{equation}\label{eq:double_decomposition2}
 \vect u = \overline{\vect u} + \vect u' \ \ \text{and} \ \  p=\overline{p}+p'.
\end{equation}
This ansatz is substituted into the forced Navier--Stokes equations \eqref{eq:gov_equations_forced} and subtracted from the mean flow equations \eqref{eq:gov_equations_meanflow}. The forcing terms are 
assumed to be constant between the time-averaged and the perturbed state and  cancel out. By neglecting the nonlinear terms $-\overline{\vect u'\cdot \nabla \vect u'}$, we obtain the perturbation equations for the mean flow
\begin{subequations}\label{eq:gov_equations_perturbation2}
\begin{align}
\frac{\partial{\vect u'}}{\partial{t}} + \vect u'\cdot\nabla\overline{\vect u} + \overline{\vect u}\cdot\nabla\vect u'&= -\nabla p' +\frac{1}{\mathrm{Re}}\nabla^2\vect u' \\
 \nabla\cdot \vect u' &= 0.
\end{align}
\end{subequations}
As noticed by \cite{Barkley2006}, these equations only apply where the forcing term $\vect F\textcolor{black}{^*}$ is constant in time to leading order. \myred{Note that the modification of the mean flow stability through the coherent Reynolds stress ($\overline{\widetilde{\vect u}\cdot \nabla \widetilde{\vect u}}$) is indirectly accounted for when solving \eqref{eq:gov_equations_perturbation2}, while any nonlinear mode-mode interaction is neglected ($-\overline{\vect u'\cdot \nabla \vect u'}=0$).} 
\subsection{Solution for \myred{the slowly diverging jet}}
Equation \eqref{eq:gov_equations_perturbation2} is first solved for a parallel flow \myred{$\overline{\vect u}_0=\left(f(r),0,0\right)^T$}. 
In describing the saturated state, our interest lies in the {\itshape spatial} growth and decay of instabilities. Therefore, the perturbations have the form 
\begin{equation}\label{eq:parallel_perturbation_ansatz}
\vect u'(\vect x,t) = \hat{\vect u}_0(r)\mathrm{e}^{i(\alpha x+m\theta-\omega t)} + {\color{black}{\it c.c.}},
\end{equation}
with complex spatial wavenumber \myred{$\alpha=\alpha_r+i\alpha_i$}, integer real azimuthal wavenumber $m$, and real temporal oscillation frequency $\omega$. The conjugate complex of the perturbation is indicated by 'c.c.'. 
The imaginary part of $\alpha$ corresponds to the spatial growth rate of the parallel flow and determines whether a perturbation of a given $m$ and $\omega$ grows ($-\alpha_i>0$) or decays ($-\alpha_i<0$) in the streamwise direction. 
Substituting  the ansatz \eqref{eq:parallel_perturbation_ansatz} and the equivalent for the pressure into \eqref{eq:gov_equations_perturbation2} leads to the eigenvalue problem 
\begin{equation}\label{eq:EVP1}
\mathbf {\color{black}{D}}(\omega)\vect \psi_0=\alpha\mathbf {\color{black}{E}}(\omega)\vect \psi_0,
\end{equation}
with the eigenvalue $\alpha$ and the eigenfunction $\vect \psi_0=(\hat{u}_0,\hat{v}_0,\hat{w}_0,\hat{p}_0)^T$, and the matrices $\mathbf {\color{black}{D}}$ and $\mathbf {\color{black}{E}}$ containing the \myred{parallel flow profiles $\overline{\vect u}_0$}. 
\par
The stability analysis is extended to weakly nonparallel flows by adopting the correction scheme developed by \cite{Crighton1976}. \myred{To account for the slow jet divergence, we} introduce a slow axial scale $X=\epsilon x$, where $\epsilon\ll 1$, and a radial component $\overline{v}_1=\overline{v}/\epsilon$. 
The global perturbation field is given as
\begin{equation}\label{eq:WKB_perturbation_ansatz}
\vect u'(X,r,\theta,t;\epsilon) = N(X) \hat{\vect u}_0(X,r;\epsilon)\exp\left(\frac{i}{\epsilon}\int^X\alpha(\xi)\mathrm{d}\xi+im\theta-i\omega t\right) + {\color{black}{\it c.c.}},
\end{equation}
with the amplitude factor $N(X)$ given \myred{as}
\begin{equation}\label{eq:GK}
 \frac{\mathrm{d}N(X)}{\mathrm{d}\color{black}X}G(X)+N(X)K(X)=0.
\end{equation}
\myred{The parameter $\epsilon$ quantifies the streamwise spreading of the jet's shear layer, with $\epsilon \sim \frac{\mathrm{d}\delta}{\mathrm{d}x}$. It is below $0.04$ throughout this study in approximate agreement with the value of $0.03$ reported by \cite{Crighton1976}.} The expressions $G$ and $K$ in \eqref{eq:GK} are derived from the solvability condition of the first order problem, and they contain the radial and streamwsie derivatives of eigenfunctions and their adjoints of the (zero order) parallel flow solution.
\par
To obtain the global perturbation field $\vect u'$ as given by \eqref{eq:WKB_perturbation_ansatz}, the \myred{eigenvalue problem \eqref{eq:EVP1} is solved for each streamwise station separately using the fitted axial velocity profile given by \eqref{eq:Velfit}.  The resulting 'local'} eigenvalues $\alpha$ and eigenfunctions $\vect \psi_0$ and adjoints are stored on disc. The eigenfunctions are  then renormalized to ensure smooth gradients in the streamwise direction and the quantities $G$ and $K$ are calculated. 
Note that the solution \eqref{eq:WKB_perturbation_ansatz} is independent of the adopted normalization of the eigenfunctions $\vect \psi_0$ as it is compensated by the amplitude factor $N$. The exact formulation of the eigenvalue problem, the derivations of $G$ and $K$, and details to the numerical scheme are given in the appendix \ref{sec:app1}.
\par
The parameter $\epsilon$ in \eqref{eq:WKB_perturbation_ansatz} is then formally dropped, and the overall shape of the instability wave excited at the nozzle lip ($x=0$) at a frequency $f$ is given by
\begin{equation}\label{eq:WKB_perturbation_ansatz2}
\vect u'(\vect x,t) = \Re\left\{\hat{\vect u}_f\mathrm{e}^{-i2\pi f t}\right\}, 
\end{equation}
with the three-dimensional shape function 
\begin{equation}\label{eq:WKB_perturbation_coefficient}
\hat{\vect u}_f(\vect x) = \hat{\vect u}_0\exp\left(i\int^x_{0}\alpha(\xi)\mathrm{d}\xi+i\frac{K}{G}+im\theta\right), 
\end{equation}
where $\hat{\vect u}_0$ and $\alpha$ represent the (zero order) parallel flow solution and depend parametrically on $x$. In this study we focus on axisymmetric perturbations and set $m=0$ throughout.

\section{Impact of forcing on the stability waves}
\label{sec:AmpVariation}
In this section, we analyze the flow's response to forcing over a wide range of amplitudes. 
The correction of the mean flow and the change of the streamwise growth of the excited instability waves are monitored. The experimental findings are compared to the stability wave model with the aim of detecting the potential limitations of the linear mean flow analysis. 
\par
\begin{figure}
\centering
\input{plots/POD_ev_A5d0_allA.tex}
\includegraphics[draft=false,width=.8\textwidth]{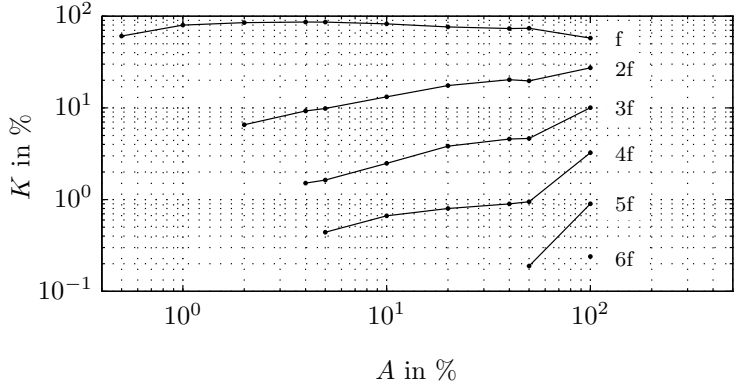}
\caption{Development of the energy contents of the fundamental and the higher order harmonics as the forcing amplitude is increased.}
\label{fig:POD_ev_A5d0_allA} 
\end{figure}

The jet is forced at amplitudes ranging from  $A=0.1~$\% to $A=100~$\%. 
The fundamental oscillation of the flow field and its higher order harmonics are derived from POD (see appendix \ref{sec:app2}). 
The overall energy content of these oscillatory modes are shown in figure \ref{fig:POD_ev_A5d0_allA} as a percentage of the total fluctuating kinetic energy. Note that the $x$- and $y$- axis are in logarithmic scale indicating that the  forcing amplitudes considered as well as the flow oscillation energy vary over multiple orders of magnitude. $A=0.5~$\% is the lowest amplitude at which the POD-based approach reveals the fundamental wave with sufficient confidence. Increasing the amplitude to $4~$\% results in an increase of the relative energy content in the fundamental, which is simply caused by an increase of the signal to noise ratio. 
The decay of the relative energy content at higher amplitudes is attributed to the redistribution of the total kinetic energy from the fundamental to the higher order harmonics. 
The first harmonic is detectable for $A\geq2~$\%. Its energy content increases linearly with $A$, reaching $30~$\% of the total energy for $A=100$\%. This is followed by higher order harmonics increasing at the same rate but at lower levels. The successive appearance of higher order harmonics with increasing forcing amplitude reveals the strong nonlinearities that are involved in the saturation of the fundamental wave.
\par
\myred{The streamwise growth of the excited instability wave is quantified by using the following amplitude measure 
\begin{equation}\label{eq:CohAmp}
\widetilde{A}(x;\hat{u}_f)=\frac{\sqrt{2}}{U_{\mathrm{jet}}D/2}\left(\int_0^{\infty}|\hat{u}_f|^2r\mathrm{d}r\right)^{1/2}.
\end{equation}
It corresponds to the amplitude of the axial velocity fluctuation at the fundamental frequency averaged across the shear layer. The corresponding complex shape function $\hat{u}_f$ can be derived from the measured phase-averaged axial velocity component (see equation \eqref{eq:shape_function}) and from the stability wave model (see equation \eqref{eq:WKB_perturbation_coefficient}).
}
\par
\begin{figure}
\centering
\input{plots/POD_amp_vs_x_delta_vs_x.tex}
\includegraphics[width=.8\textwidth]{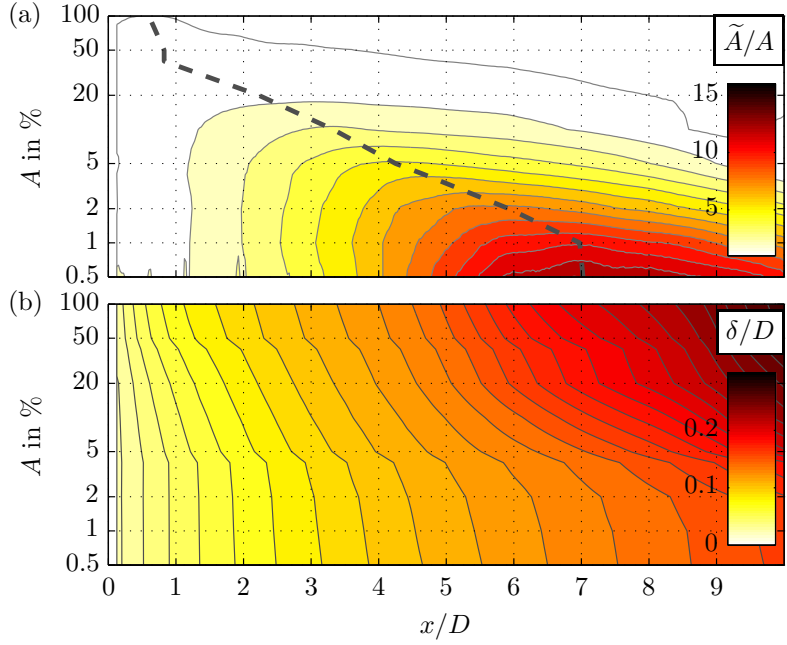}
\caption{(a) Coherent fundamental amplitude $\widetilde{A}_f$ of the axial velocity component normalized by the forcing amplitude $A$ representing the gain of the incoming perturbations as a function of streamwise distance and forcing amplitude.
The dashed line marks the streamwise location of maximum gain.
(b) Momentum thickness of the mean flow as a function of streamwise distance and forcing amplitude.}
\label{fig:POD_amp_vs_x_delta_vs_x} 
\end{figure}

\myred{Figure \ref{fig:POD_amp_vs_x_delta_vs_x}a shows $\widetilde{A}$ derived from the measurements as a function of the forcing amplitude and streamwise location. Quantities are normalized with respect to the forcing amplitude $A$ and represent the gain of the inlet perturbations.} Note that the $y$-axis is again shown in a logarithmic scale to magnify the low amplitude forcing regime. 
Figure \ref{fig:POD_amp_vs_x_delta_vs_x}b shows contours of the corresponding momentum thickness $\delta$, indicating the change of the mean flow with increasing 
forcing amplitude. 
\par
At very low forcing amplitudes, the amplification in the shear layer is strong and the oscillations are amplified by a factor of $15$. Upon increasing the forcing amplitude, the gain in the shear layer decreases continuously and converges to a level only slightly above $1$ for $A\geq50~$\%. Simultaneously, the streamwise location of maximum gain, indicated by a dashed line in figure \ref{fig:POD_amp_vs_x_delta_vs_x}a, is shifted upstream to the vicinity of the nozzle lip. Noting the logarithmic scale of the $y$-axis, this shift increases exponentially with $A$, demonstrating the high sensitivity of the growth rates to changes in the forcing amplitude. 
Similarly, the downstream growth of the shear layer is significantly enhanced with increasing forcing, as indicated by the increase of $\delta$ shown in figure \ref{fig:POD_amp_vs_x_delta_vs_x}b. 
\par
\begin{figure}
\centering
\input{plots/LSA_alpha_contour_allA.tex}
\includegraphics[width=1\textwidth]{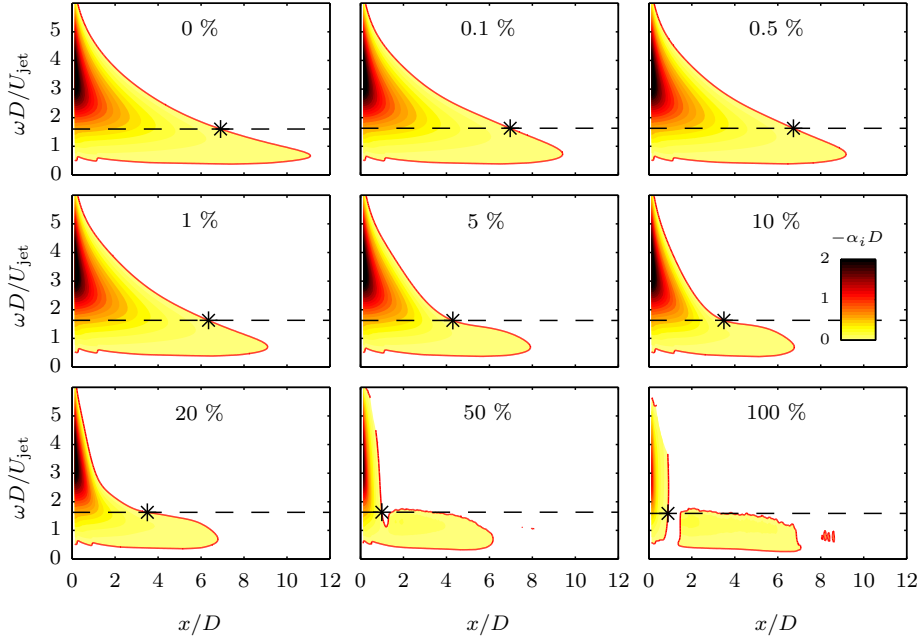}
\caption{Contours of streamwise growth rate derived from the quasi parallel flow analysis as a function of frequency and streamwise station. The thick red line refers to the neutral curve of instability, the dashed horizontal line indicates the forcing frequency, and the black star marks the neutral point of the forced instability wave. Modification of the mean flow \myred{stability} with higher forcing amplitude is clearly visible by the distortion of the neutral curve.}
\label{fig:LSA_alpha_contour_allA} 
\end{figure}

 
Figure \ref{fig:LSA_alpha_contour_allA} depicts the \myred{theoretical results showing the} growth rate of the parallel flow analysis, $-\alpha_i$, computed for a wide range of frequencies and streamwise locations. These stability maps provide an overview of how the forcing affects the mean flow stability. The dashed line indicates the forcing frequency selected in this study. As can be seen from the top-left plot, the forcing frequency corresponds to a wave that, under natural conditions, undergoes significant growth near the nozzle and saturates at approximately $x/D=6$, where the neutral curve intersects with the dashed horizontal line.  Waves at higher frequencies undergo stronger amplification near the nozzle, but saturate at an earlier streamwise location, whereas the opposite applies for lower frequencies. At a downstream distance of $x/D\approx11$ all frequencies are damped, rendering the jet's far-field stable to axisymmetric perturbations. 
\par
The weakest forcing considered ($A=0.1~$\%) is already sufficiently strong to noticeably alter the stability map. The neutral point of the low frequency waves is shifted considerably upstream. \textcolor{black}{Apparently, at such low forcing amplitudes, the excited waves are still strong enough to alter the mean flow stabiliy, although they are not detected in the POD analysis.}
\par  
Upon further increasing the forcing amplitude, the neutral curve is significantly distorted. The neutral point at the forcing frequency is shifted upstream until it reaches the proximity of the nozzle. The growth rates at the nozzle remain unaffected over a wide range of forcing amplitudes, but ultimately decrease for the very strong forcing. 
The stability map for the jet forced at $A>50~$\% features two separate regions of spatial growth. As discussed later, the amplification in the \textcolor{black}{downstream region of spatial amplification} is not confirmed by the experiments and is not considered as physical, rendering the strongly forced jet unstable in a very small streamwise extent. 
\par
\begin{figure}
\centering
\input{plots/LSA_POD_xNeutral.tex}
\includegraphics[width=.8\textwidth]{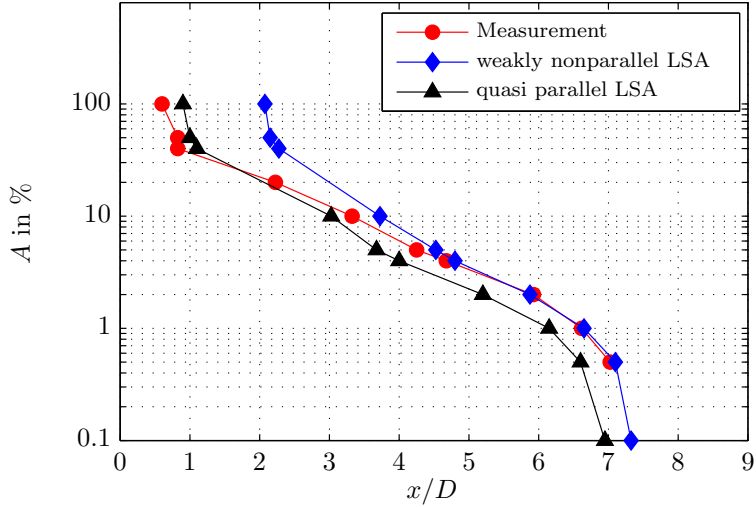}
\caption{Streamwise station of neutral amplification \myred{(neutral point)} as a function of forcing amplitude as derived from the weakly nonparallel flow analysis, the parallel flow analysis, and the measurements.}
\label{fig:LSA_POD_xNeutral} 
\end{figure}

The \myred{streamwise location of neutral stability (neutral point)} is of particular interest. \myred{There}, the instability wavetrain has gone through its entire amplification cycle and reaches its maximum amplitude. For the parallel flow analysis, neutral \myred{instability} corresponds to $\alpha_i=0$, while for the weakly nonparallel flow analysis, we define neutral \myred{instability} as \textcolor{black}{$\mathrm{d}\widetilde{A}/\mathrm{d}x=0$}. Figure \ref{fig:LSA_POD_xNeutral} depicts \myred{the neutral point} as a function of the forcing amplitude together with the experimental findings. 
The weakly nonparallel flow solution is consistent with the experiments for forcing amplitudes lower than $5~$\%, while the quasi parallel flow solution slightly underestimates the amplified regime. With stronger forcing, the previous starts to deviate from the experimental results, while the latter compares reasonably well with the experimental data.
\par
\begin{figure}
\centering
\input{plots/LSA_POD_phase_amp_profiles_allA.tex}
\includegraphics[width=1\textwidth]{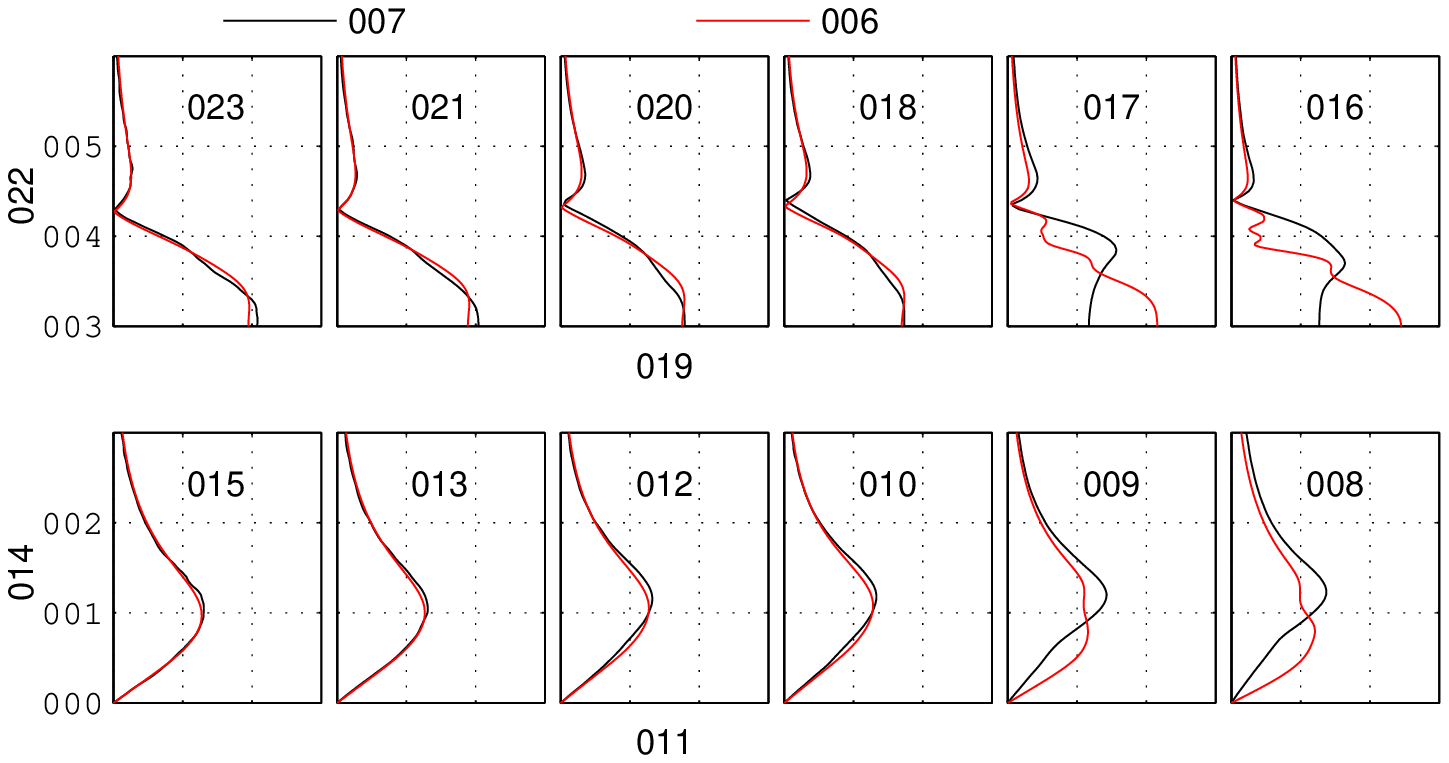}
\caption{Amplitude of the shape function at the neutral point. LSA prediction deviates from the measurements at high forcing, as the neutral point is located close to the nozzle and the stability wave interacts with the nozzle lip. \myred{Amplitudes are normalized with respect to the area under the graph}}
\label{fig:LSA_POD_phase_amp_profiles_allA} 
\end{figure}

The inaccuracy of the weakly nonparallel flow analysis at strong forcing (and stronger nonparallelism), is not intuitive and requires further investigation. 
Figure \myred{\ref{fig:LSA_POD_phase_amp_profiles_allA}} shows a comparison of the local eigenfunctions with the measurements at the neutral point. 
These quantities are readily given by the eigenfunction of the parallel flow solution $\hat{\vect u}_0$ and do not depend on the weakly nonparallel flow approximation. 
The prediction is excellent for weakly and moderate forcing amplitudes, whereas discrepancies in the inner jet region are noticeable for very strong forcing.
The reason for the discrepancy is an interaction of the excited instability wave with the nozzle that is not incorporated in the model. As shown by \cite{Orszag1970} and \cite{Rienstra1983}, these interactions are significant within a downstream distance of half a wavelength of the growing instability wave, which corresponds to  $x \approx 1.8D$ for the present case.
This becomes a serious issue once the neutral point of the instability shifts close to the nozzle lip, which is the case at very high amplitude forcing.
Then, the local eigenfunctions deviate significantly from the measurements, which induces significant errors in the weakly nonparallel flow correction scheme, and the neutral point is predicted more accurately by the quasi parallel flow solution. 
\section{Detailed comparison of the stability wave model with experiments}
\label{sec:detailed_comparison}
The stability wave model predicts the point of neutral stability for a wide range of forcing amplitudes up to values where the flow is essentially stable and waves decay shortly downstream of the nozzle. 
Discrepancies arise for very strong forcing, where the small growth that occurs close to the nozzle is affected by the geometric confinement. 
In the following, we focus on the flow forced at moderate amplitudes ($A=5~$\%) and undertake a point-wise comparison of the instability wave model with the perturbation field reconstructed from the PIV measurements.
\subsection{Overall mode shape characteristics}
\begin{figure}
\centering
\input{plots/LSA_POD_contour_A5d0.tex}
\includegraphics[width=\textwidth]{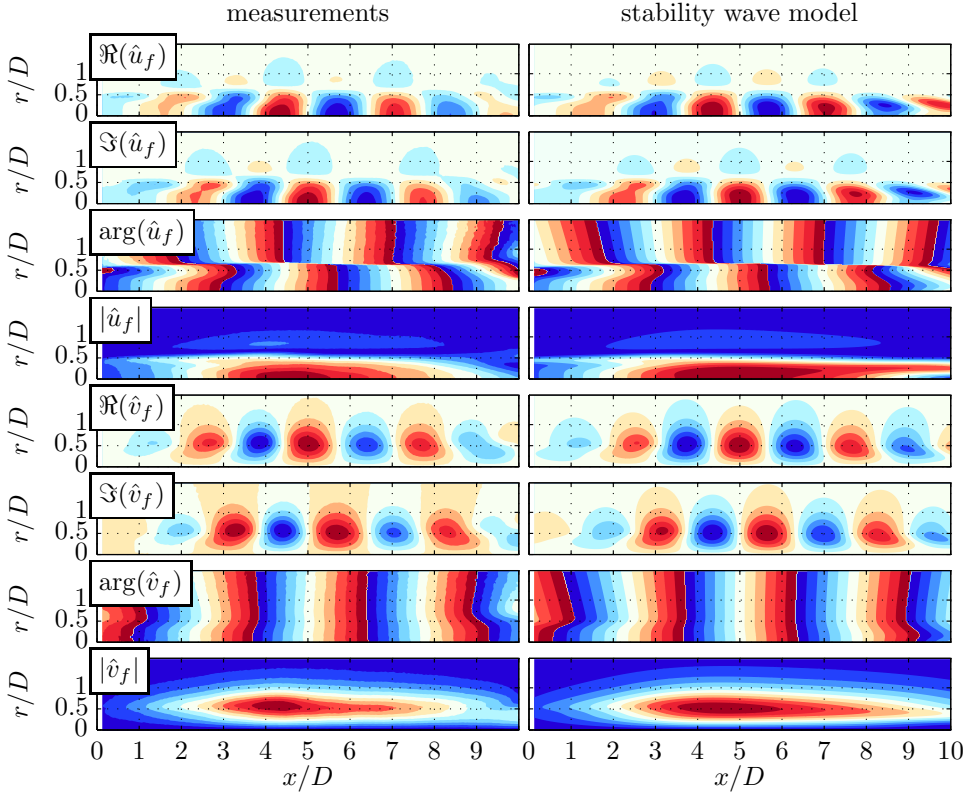}
\caption{Jet forced at $A=5~$\%: Overall comparison of the flow oscillations measured via PIV (left) and the perturbation field obtained from the LSA (right). Displayed are the real and imaginary part, phase, and magnitude of the streamwise and radial component of the velocity fluctuations. \myred{Real and imaginary part are shown at an arbitrary phase reference.}}
\label{fig:LSA_POD_contour_A5d0} 
\end{figure}

Figure \ref{fig:LSA_POD_contour_A5d0} provides an overview of the mode shape obtained from the  PIV and the LSA.
The experimental results are presented in the left column showing the real and imaginary part, amplitude, and phase of the coherent velocity oscillating at the fundamental frequency. These quantities are exclusively given by the first two POD modes (see \eqref{eq:POD_reconstruction} in appendix \ref{sec:app2} for details). The right column displays the weakly nonparallel flow solution computed for the fundamental frequency from \eqref{eq:WKB_perturbation_ansatz2}. 
The agreement is excellent in the upstream half of the measurement domain, which corresponds approximately to the growth region of the wavetrain. 
Further downstream, discrepancies in the phase and amplitude distribution, particularly for the streamwise component, become noticeable. The real and imaginary part of the velocity fluctuations represent the instantaneous wave shape with a phase lag of $\pi/2$. Together they represent the downstream propagation of the forced instability wave. 
The shape of the axial and radial component tilt from forward leaning in the growth region to backward leaning in the decaying region. This indicates the variation of the phase velocity in the radial direction, with high velocities near the jet centerline and low velocities in the shear layer. This is very well captured by the stability wave model. 
\subsection{Radial phase and amplitude distribution}
\begin{figure}
\centering
\input{plots/LSA_POD_phase_amp_profiles_A5d0.tex}
\includegraphics[width=\textwidth]{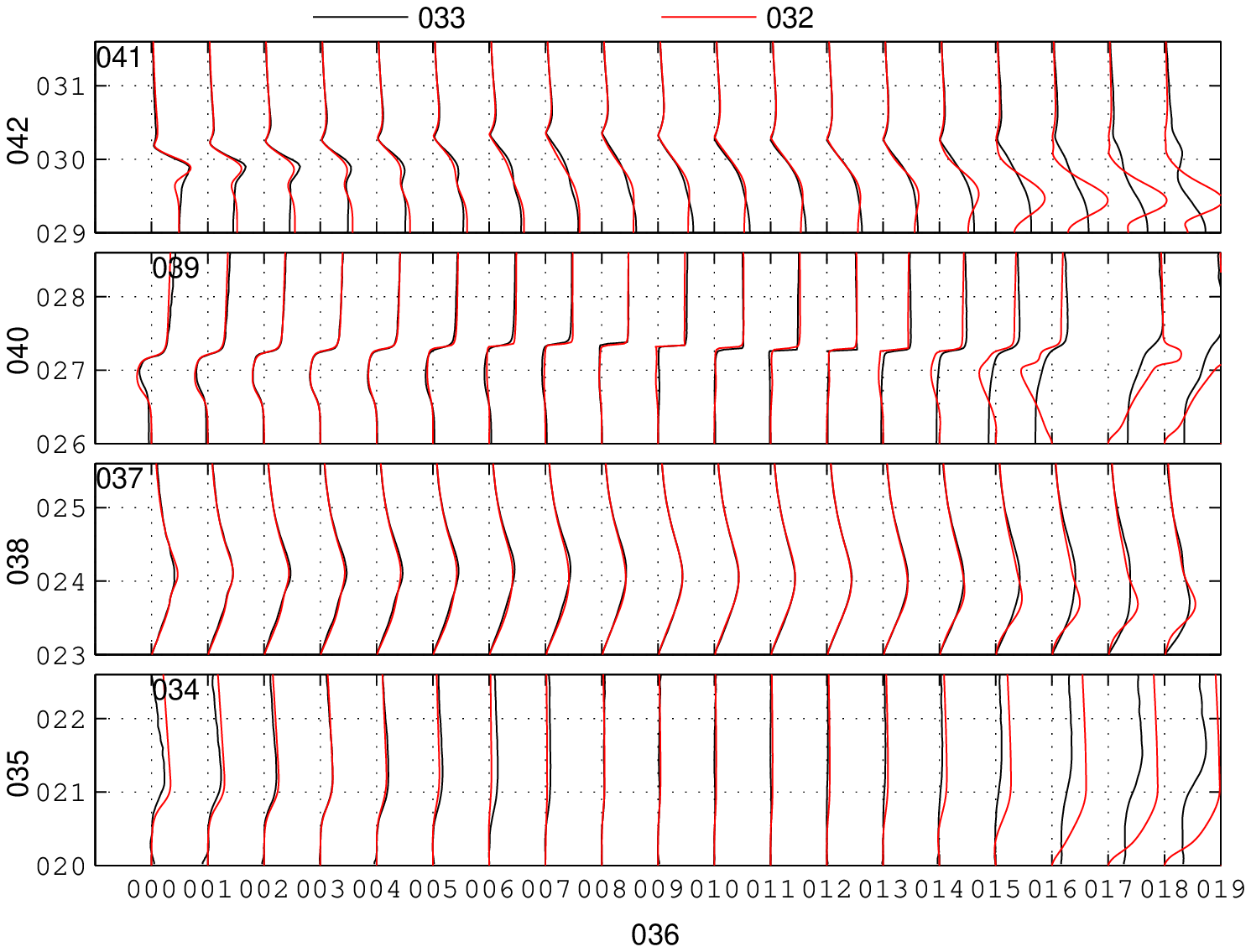}
\caption{Jet forced at $A=5~$\%: Detailed comparison of the flow oscillations measured via PIV (black solid lines) and the perturbation field obtained from the LSA (red solid lines). Displayed are the radial profiles of streamwise velocity amplitude (first row) and phase (second row), and radial profiles of radial velocity amplitude (third row) and phase (fourth row). Discrepancies are noticeable near the nozzle and far downstream in the decaying region of the wave. \myred{Amplitudes are normalized with respect to the area under the graph. Phase distributions are aligned at the radial point of maximum amplitude.}}
\label{fig:LSA_POD_phase_amp_profiles_A5d0} 
\end{figure}

The accuracy of the stability wave model is further quantified by comparing the radial amplitude and phase distribution at each streamwise location with the measurements. 
Figure \ref{fig:LSA_POD_phase_amp_profiles_A5d0} shows profiles of the phase and magnitude for several streamwise locations for the streamwise and radial component, respectively. 
Discrepancies are noticeable within the first two nozzle diameters, indicating the regime where the instability wave interacts with the nozzle lip. In the region where the wave reaches its maximum amplitude ($x/D\approx4$), the LSA predictions are nearly indistinguishable from the measurements, while in the decaying region of the wave, the predictions start to deviate from the measurements at approximately $x/D=7.5$. 
\subsection{Streamwise growth rate and phase velocity}
A demanding test for the accuracy of the stability wave model is the comparison of the predicted growth rates with the measurements.
For the weakly nonparallel flow solution \eqref{eq:WKB_perturbation_ansatz}, the streamwise growth rate and phase velocity is not only determined by the exponent of the parallel flow solution, ($i\alpha-\omega t$), but also depends on the streamwise change of the local eigenfunctions $\hat{\vect u}_0$. 
For consistent comparison with experimental data, we define the streamwise growth rate as 
\begin{equation}\label{eq:def_streamwise_growthrate}
\widetilde{\textcolor{black}{\alpha}}_i(x) := -\frac{\mathrm{d}(\ln\textcolor{black}{\widetilde{A}})}{\mathrm{d}x}, 
\end{equation}
\myred{where the amplitude $\widetilde{A}$ is calculated from the axial velocity component as defined by \eqref{eq:CohAmp}, and we define the} streamwise phase velocity as 
\begin{equation}\label{eq:def_streamwise_phasevelocity}
\widetilde{\textcolor{black}{c}}_{\mathrm{ph}}(x,r) := \frac{\omega}{\partial \textcolor{black}{\varphi}/\partial x} \ \ \ \text{with} \ \ \ \textcolor{black}{\varphi}(x,r) = \arg(\textcolor{black}{\hat{u}_f}).  
\end{equation}
\par
\begin{figure}
\centering
\input{plots/LSA_POD_cph_growthrate_vs_x_A5d0.tex}
\includegraphics[draft=false,width=.8\textwidth]{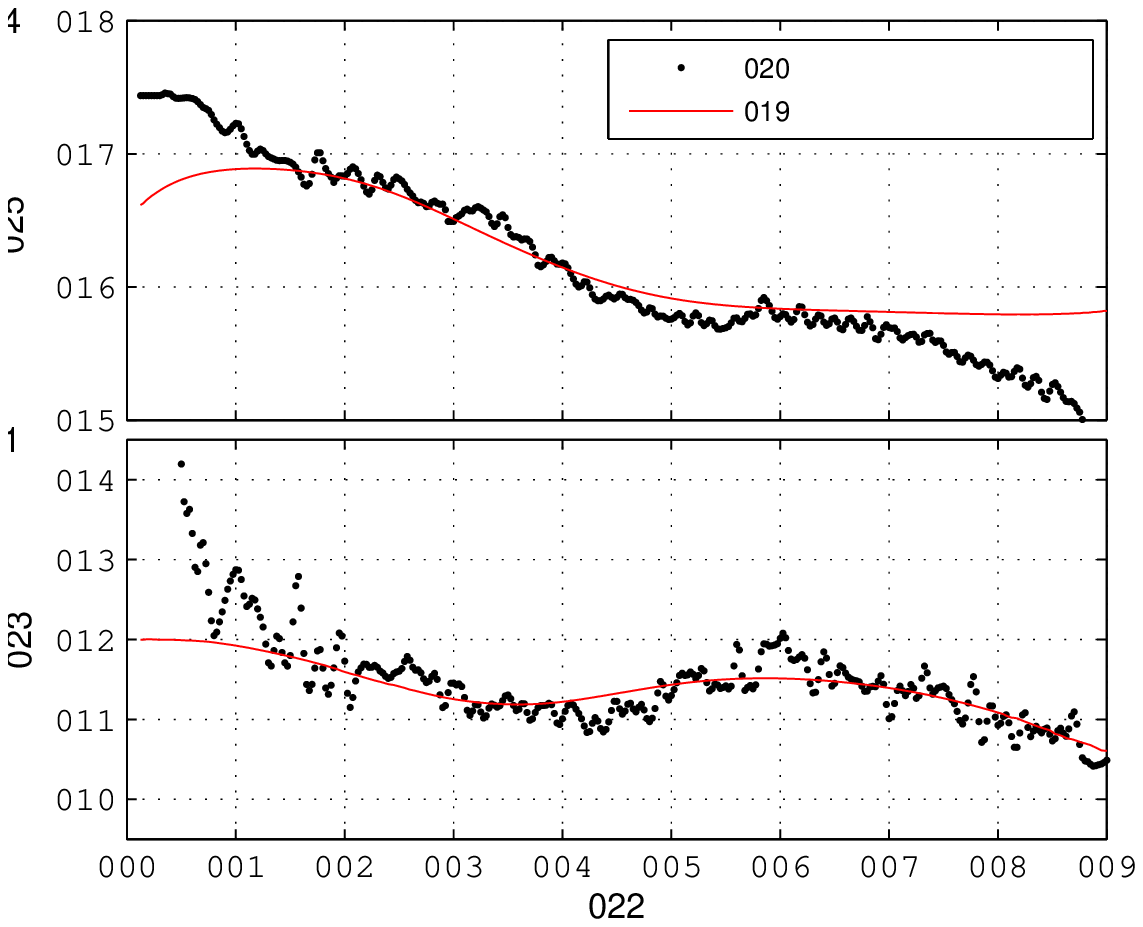}
\caption{Jet forced at $A=5~$\%: (a) Streamwise distribution of streamwise growth rate as defined by \eqref{eq:def_streamwise_growthrate}. (b) Streamwise distribution of streamwise phase velocity as defined by \eqref{eq:def_streamwise_phasevelocity}. Phase velocities are extracted at the half velocity radius $R_{.5}$. Black dots refer to measurements of the forced fundamental wave, red solid line refers to the weakly nonparallel flow solution.}
\label{fig:LSA_POD_cph_growthrate_vs_x_A5d0} 
\end{figure}

Figure \ref{fig:LSA_POD_cph_growthrate_vs_x_A5d0}a shows the growth rates derived from the stability wave model together with the growth rates of the measured fundamental wave. 
The growth rates continuously decay in the downstream direction and cross zero at approximately $x/D=4$. In the vicinity of the nozzle, the growth rates are slightly underestimated. 
Downstream of $x/D=2$, the wave model captures the growth rates remarkably well up to a downstream distance of $x/D=7$.
These results are in line with the previously discussed comparison of the local eigenfunctions, which suggests that the inaccurate growth rates stem from the deteriorated prediction  of the local eigenfunctions near the nozzle and far downstream. However, it should be noted here, that the growth rates derived from the measurements scatter near the nozzle and in the downstream part where the amplitude of the instability wave and, consequently, the signal to noise level is low. 
\par
Figure \ref{fig:LSA_POD_cph_growthrate_vs_x_A5d0}b shows the phase velocity extracted at the half velocity radius $R_{.5}$.  
In contrast to the growth rates shown previously, the phase velocity given by \eqref{eq:def_streamwise_phasevelocity} does not represent radial integrals and, therefore, scatters even more near the nozzle. Nonetheless, figure \ref{fig:LSA_POD_cph_growthrate_vs_x_A5d0}b clearly shows that the wave model allows for an excellent prediction of the phase velocity in the growing and decaying region of the perturbation, except for the region close to the nozzle.

\section{Mean-coherent and mean-incoherent energy transfer}
\label{sec:interaction}
The previous section revealed two regions where the stability wave model deviates from the measurements.
The discrepancies near the nozzle are readily explained by an interaction of the instability wave with the geometric confinement, but what causes the significant lack of comparison of the local eigenfunctions in the decaying region of the instability wave? 
\par
\begin{figure}
\centering
\input{plots/PIV_amp_prod_versus_x_A5d0_modified.tex}
\includegraphics[height=.35\textheight,trim=2.5cm -.2cm 0 0]{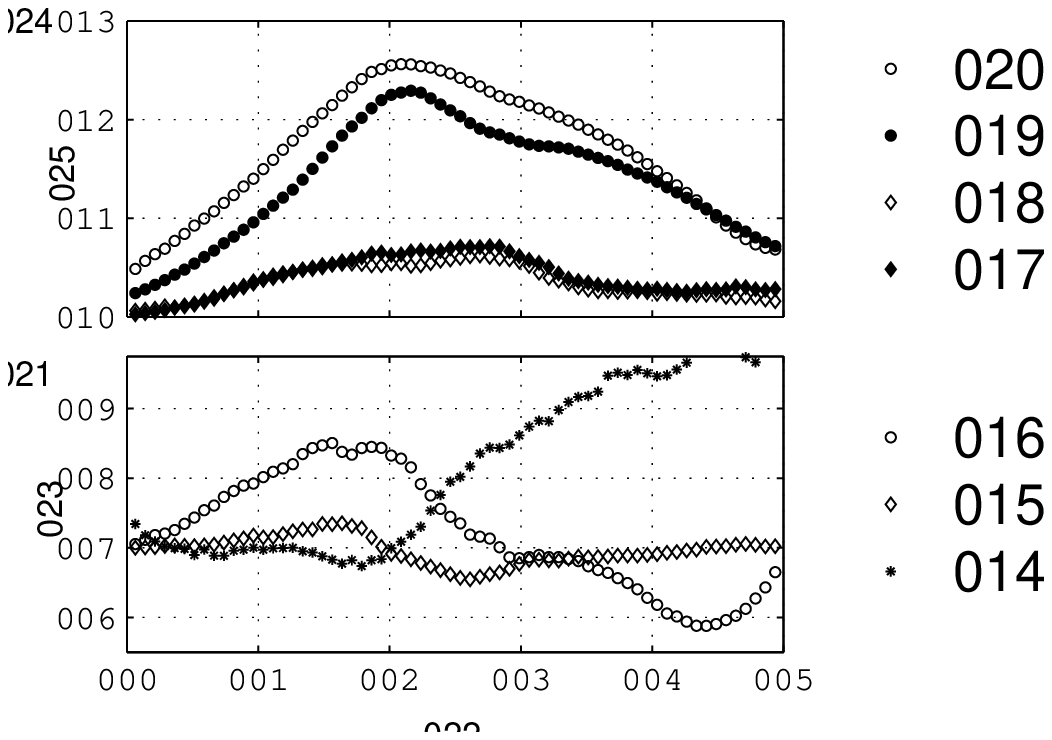}
\caption{The jet forced at $A=5~$\%: (a) Streamwise amplitude distribution of the fundamental and first harmonic for the streamwise and radial velocity component, respectively; (b) Production terms for the fundamental, first harmonic, and uncorrelated fluctuations. All quantities are obtained from PIV measurements}
\label{fig:PIV_amp_prod_versus_x_A5d0} 
\end{figure}

In order to adress this question, we investigate the energy flux between the mean flow, the fundamental, and the higher order harmonics during the growth and decay phase of the instability wave. 
Figure \ref{fig:PIV_amp_prod_versus_x_A5d0}a shows the streamwise amplitude distribution of the fundamental and the first harmonic for the streamwise and radial velocity component, respectively, for the jet forced at $A=5~$\%. 
Similarly to figure \ref{fig:LSA_POD_phase_amp_profiles_allA}a, the amplitudes are normalized by the perturbations at the nozzle exit plane given by the piston oscillating amplitude $A$. For the axial component of the fundamental wave, it represents the gain, and must be unity at the nozzle exit. 
For the particular forcing amplitude of $A=5~$\%, the overall gain in the streamwise component of the fundamental wave is approximately $5$. 
The fundamental streamwise and radial oscillations saturate at roughly the same streamwise location, which corresponds to the region where the shear layer roll-up is completed. 
The amplitude of the first harmonic grows significantly in the downstream direction, saturating further downstream than the fundamental.  
The overall gain of the streamwise component of the first harmonic is approximately $25$, which corresponds to the square of the gain of the fundamental. This strongly indicates that the first harmonic is caused by nonlinearities of the fundamental wave and not by higher order harmonics accidentally introduced by the apparatus. Therefore, the oscillations at the higher order harmonics cannot be predicted from the stability wave model that assumes a forcing localized at the nozzle. 
\par
The concept of the triple decomposition allows the exploitation of the interaction between the mean flow, the wave-induced oscillations, and the uncorrelated oscillations. 
Neglecting viscous dissipation for the moment, the energy equation for the mean flow can be written in the following form \citep{Reynolds1972}
\begin{equation}\label{eq:dEdx}
 \frac{1}{2}\frac{\mathrm{d}}{\mathrm{d}x}\int_{r=0}^{\infty}\overline{u}^3r\mathrm{d}r = -\int_{r=0}^{\infty}-\overline{\widetilde{u}\widetilde{v}}\frac{\partial\overline{u}}{\partial r}r\mathrm{d}r-\int_{r=0}^{\infty}-\overline{u''v''}\frac{\partial\overline{u}}{\partial r}r\mathrm{d}r.
\end{equation}
The first and second term on the right side of \eqref{eq:dEdx} represent the energy transferred between the mean flow and the coherent motion and the mean flow and the uncorrelated motion, respectively, representing the work of the Reynolds stresses against the shear of the axial velocity component. The coherent energy production term can be further decomposed into its fundamental and higher order harmonics  
$$
\int_{r=0}^{\infty}\overline{\widetilde{u}\widetilde{v}}\frac{\partial\overline{u}}{\partial r}r\mathrm{d}r = \int_{r=0}^{\infty}\overline{\widetilde{u}_f\widetilde{v}_f}\frac{\partial\overline{u}}{\partial r}r\mathrm{d}r+
\int_{r=0}^{\infty}\overline{\widetilde{u}_{2f}\widetilde{v}_{2f}}\frac{\partial\overline{u}}{\partial r}r\mathrm{d}r+
\int_{r=0}^{\infty}\overline{\widetilde{u}_{3f}\widetilde{v}_{3f}}\frac{\partial\overline{u}}{\partial r}r\mathrm{d}r+...,
$$
since all cross-frequency products, such as e.g.  $\overline{\widetilde{u}_f\widetilde{v}_{2f}}$, vanish. 
\par
The production terms derived from the measured oscillating flow field are shown in figure \ref{fig:PIV_amp_prod_versus_x_A5d0}b for the fundamental, the first harmonic, and the non-coherent part. 
In the growth region of the fundamental wave ($x/D<4$), the flow is essentially laminar and dominated by one oscillatory mode and its harmonics. 
Production of the fundamental and the first harmonic are positive, indicating energy flux from the mean flow to the oscillating flow. The Reynolds stresses of the uncorrelated fluctuations are essentially zero. 
The fundamental amplitude saturates at $x/D\approx 4$ where it still receives energy from the mean flow. The excess energy is partly transferred to the first higher harmonic
 and to the uncorrelated fluctuations.  The first harmonic transfers this energy back to the mean flow (see corresponding negative production), to higher order harmonics, and to the uncorrelated fluctuations. The production term of the latter increases continuously during the decay of the fundamental indicating significant energy flux from the mean flow to fluctuations that are not phase-locked with the forcing. At $x/D>6$ the production of the fundamental wave becomes negative and energy of the coherent motion is transferred back to the mean flow. 
A reversed energy flow cascade from smaller scales (coherent motion) to larger scales (mean flow) was also observed in the forced mixing layer where the mixing layer thickness was, in fact, observed to decrease in the region of negative production \citep{Weisbrot1988}. \textcolor {black}{Consistent with these previous investigations, the negative production term is caused by a change of sign of the Reynolds stress of the forced coherent structure and not by a distortion of the mean flow.}
\par
Based on the fact that the stability wave model deviates from the measurements downstream of $x/D>7$ (see e.g. figure \ref{fig:LSA_POD_phase_amp_profiles_A5d0}), we propose the following hypothesis:
The stability analysis based on the mean flow can predict the fundamental wave oscillation as long as it gains energy from the mean flow. 
This is the case during its entire growth phase and during part of the decay phase, where the excess energy is drained by the higher order harmonics, other modes, and the uncorrelated fluctuating field. Once the mean-coherent energy flux is reversed, the coherent fluctuations are no longer linearly  determined by the mean flow and the adopted wave model breaks down. 
\section{Summary and conclusions}
\label{sec:conclusions}
A stability wave model is developed to predict the convective vortex pattern observed in open shear flows.  
The model is based on a local linear spatial stability analysis with a correction for weakly nonparallel flows. 
The adopted scheme allows for consistent normalization of the parallel flow solutions to assemble the overall perturbation field.
The stability analysis is applied to the time-mean flow inherently taking the mean flow modifications induced by the finite amplitude perturbations into account. 
The evolution equations linearized around the mean describe the growth, saturation, and decay of the excited wavetrain. 
Mode-mode interactions as well as interactions of the instability wave with small-scale turbulence are ignored. 
\par
An experimental study is conducted to validate the performance of the mean flow stability model.
An axisymmetric laminar jet is generated by a piston-cylinder-type arrangement and released into a large water tank.  
The jet's axisymmetric mode is excited by imposing a sinusoidal motion onto the piston's mean motion at relative amplitudes ranging from $0.1~$\% to $100~$\%.
The flow field is measured using high-spatial-resolution PIV. The oscillations corresponding to the fundamental and higher order harmonics of the forcing frequency are extracted from the uncorrelated PIV snapshots using a POD-based phase-reconstruction scheme. At low amplitude forcing ($<1~$\%),
the excited instability waves undergo significant gain, saturating at around six nozzle diameters downstream of the nozzle.
The energy of the higher harmonics is orders of magnitude lower than the fundamental. 
By increasing the forcing amplitude, the gain of the fundamental wave decreases significantly and the point of saturation moves closer to the nozzle. 
Higher order harmonics become successively more energetic relative to the fundamental, which goes hand in hand with a significant mean flow modification. 
Calculations of the coherent production terms reveal that during the growth phase the fundamental wave drains energy from the mean flow, while in the decay phase the fundamental transfers energy to its higher order harmonics and to the uncorrelated fluctuations, and at a subsequent stage, transfers energy back to the mean flow. The backscatter of energy to the mean flow is remarkable and supports earlier observations in the forced mixing layer \citep{Weisbrot1988}. 
At forcing amplitudes higher than $40~$\%, the overall gain of the fundamental approaches unity and saturation occurs close to the nozzle exit. 
\par
The experimental results are compared to the stability wave model, including amplitude and phase distribution, growth rates, and phase velocities.
The overall agreement is excellent for small and moderate amplitudes.  
The stability wave model captures the growth phase reasonably well and precisely predicts the neutral point and its upstream displacement with increasing forcing amplitude. Nonlinear interactions, which are indicated by the higher order harmonics, do not seem to affect the accuracy of the prediction. In fact, the prediction is most accurate around the neutral point, where the first harmonic is most energetic relative to the fundamental. 
In close vicinity to the nozzle, discrepancies in the shape function and an underestimation of the growth rates are noticeable, which is explained by an interaction of the forced instability wave with the confinement of the nozzle, which is not accounted for in the present model.  
The decay phase of the stability wave is well predicted up to a streamwise location where the Reynolds stresses of the fundamental wave change their sign. Energy is then transferred from the coherent fluctuations back to the mean flow. The backscatter of energy from smaller to larger scales is typically a stochastic process that is not proportional to the mean, and hence, it appears plausible that the mean flow model fails at this point. 
\par
The investigation of the flow forced over a  wide range of amplitudes reveals that the stability wave model works well regardless of the nonlinearity involved. Referring back to the derivations of the mean flow stability equations, it appears that neglecting of mode-mode and coherent-turbulent interactions is less significant than one would expect by considering their relative magnitudes.
Inaccuracies of the model stemming from the interaction of the mean flow with higher harmonics, as suggested by \citep{Sipp2010a}, have not been observed in the present study. 
Yet, the pronounced deviation of the mean flow solution from the measurements in the region of energy backscatter suggests mean-coherent interactions that cannot be determined using the linear model. 
\par
\myred{In conclusion, mean flow stability analysis provides accurate prediction of growth rates and shape functions of the coherent flow oscillation regardless of the fluctuation amplitude, if the driving stability mechanisms is captured by the model. 
This gives credibility to mean flow stability analyses conducted recently, such as on the cylinder wake vortex shedding \citep{Barkley2006,Meliga2012a}, on the vortex breakdown bubble \citep{Oberleithner2011a}, and on turbulent jets \citep{Gudmundsson2011,Oberleithner2013c}.}

\section*{Acknowledgments}
The authors kindly acknowledge the stimulating discussions with Vassili Kitsios. 
The first author was supported by a fellowship within the Postdoc-Program of the German Academic Exchange Service (DAAD).
The support of the Australian Research Council (ARC) and the German Research Foundation (DFG) is greatfully acknowledged.
\appendix
\section{Phase-reconstruction scheme}
\label{sec:app2}
In continuation of earlier work \citep{Oberleithner2011a}, we utilize Proper Orthogonal Decomposition (POD) to extract the phase-averaged flow field from the uncorrelated PIV snapshots. 
This is demonstrated here for the jet forced at $A=5~$\%. 
\par
POD allows for an efficient characterization of the flow dynamics in a orthogonal subspace that is optimal in terms of the captured kinetic energy \citep{Berkooz1993}.
This involves a projection of the PIV snapshots taken at $N$ uncorrelated points at times $t_j, j=1,...,N$ on a $N$-dimensional orthogonal vector base that maximizes the kinetic energy content for any $I$-dimensional subset of the base. In other words, the POD modes provide a least--order expansion of the fluctuating flow field, that is
\begin{equation}\label{eq:PODexpantion}
 \widetilde{\vect u}(\vect x,t_j) = \sum_{i=1}^Ia_i(t_j)\vect\Phi_i(\vect x)+\vect u_{\mathrm{res}},
\end{equation}
by minimizing the residual $\vect u_{\mathrm{res}}$. The $a_i(t_j)$ is the $i$th POD coefficient corresponding to the $i$th POD mode $\vect\Phi_i$. 
The POD modes are derived from the PIV data using the snapshot method \citep{Sirovich1987}. 
The corresponding algorithm is based on a $N\times N$ autocorrelation matrix $\mathbf R=R_{kl}$ defined as 
\begin{equation}\label{eq:R_kl}
R_{kl} = \frac{1}{N}\langle \widetilde{\vect u}(\vect x,t_k),\widetilde{\vect v}(\vect x,t_l)\rangle
\end{equation}
with $\langle\widetilde{\vect u},\widetilde{\vect v}\rangle$ being the inner product between the vectors $\widetilde{\vect u}$ and $\widetilde{\vect v}$. 
To improve the signal-to-noise ratio, the autocorrelation matrix is computed for the transversal velocity fluctuation $\widetilde{v}$ only. Moreover, symmetry is enforced to the instantaneous velocity field by applying $\widetilde{v}_{\mathrm{symm}}(x,y,t_j) = (\widetilde{v}(x,y,t_j)+\widetilde{v}(x,-y,t_j))/2$. This is necessary to filter out any asymmetric mode fluctuations that become correlated to the forcing at the downstream end of the potential core of the jet.  
The autocorrelation is symmetric and positive semi-definite, and the eigenvalue problem 
\begin{equation}\label{eq:POD_EVP}
\mathbf R\vect a_i = \lambda_i\vect a_i
\end{equation}
allows for efficiently computing the POD coefficients $\vect a_i := [a_i(t_1),...,a_i(t_N)]$. 
The spatial POD modes are calculated as a linear combination of the fluctuation snapshots 
\begin{equation}
\vect\Phi_i(\vect x) = \frac{1}{N\lambda_i}\sum_{j=1}^Na_i(t_j)\widetilde{\vect u}(\vect x,t_j). 
\end{equation}
The amount of kinetic energy contained in each mode is related to the eigenvalues as
\begin{equation}\label{eq:POD_energy}
K_i:=\overline{\langle \widetilde{\vect u},\vect\Phi_i\rangle^2}/2 =\overline{a_i^2}/2=\lambda_i/2.
\end{equation}
\par
\begin{figure}
\centering
\input{plots/POD_ev_A5d0.tex}
\includegraphics[width=0.5\textwidth]{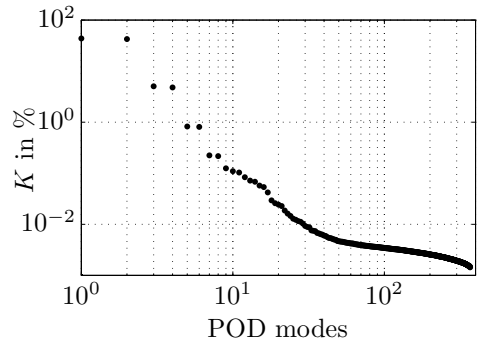}
\caption{Relative energy content of the POD modes of the jet forced at $A=5~$\%; the fundamental and the three higher order harmonics appear as mode pairs of equal energy.}
\label{fig:POD_ev_A5d0} 
\end{figure}
\begin{figure}
\centering
\input{plots/POD_modes_A5d0.tex}
\includegraphics[width=\textwidth]{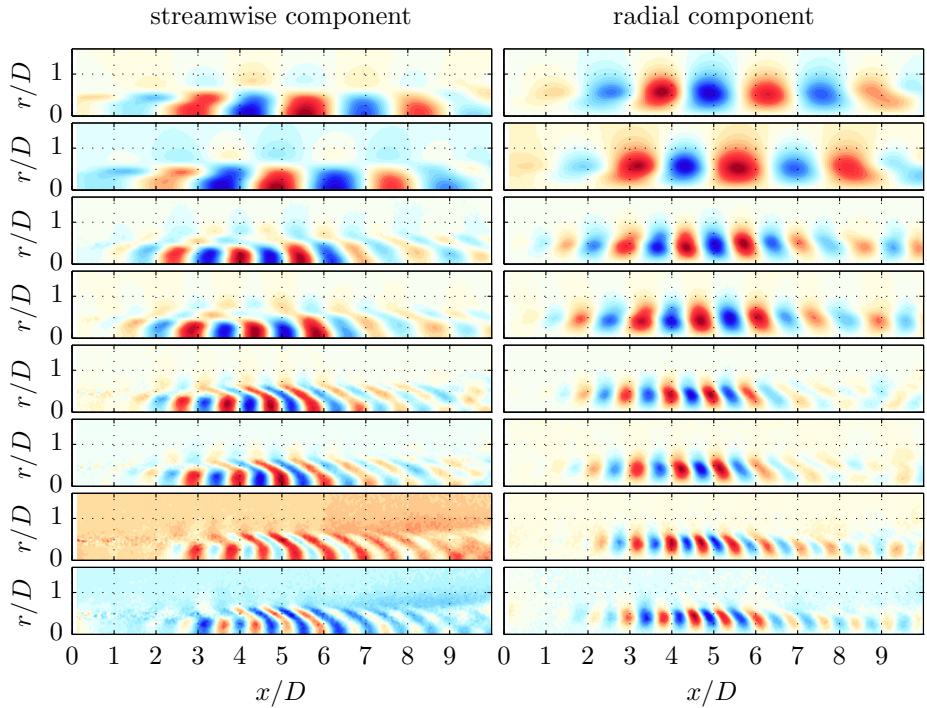}
\caption{POD modes of the jet forced at $A=5~$\%: Spatial structure of the first (top) to the eighth (bottom) POD mode indicating the fundamental and higher order harmonics. Streamwise and radial component are shown on the left and right side, respectively.}
\label{fig:POD_modes_A5d0} 
\end{figure}
\begin{figure}
\centering
\input{plots/POD_lisajour_A5d0.tex}
\includegraphics[width=\textwidth,trim=1cm 0cm 1cm 0cm,clip]{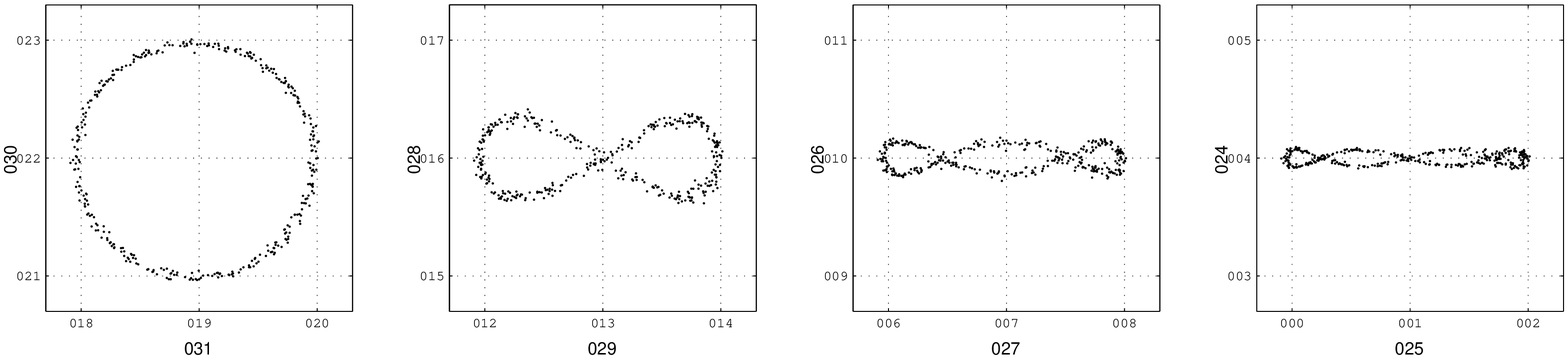}
\caption{Scatter plot of the POD coefficients of the jet forced at $A=5~$\%; The phase portrait of the fundamental (left) and higher order harmonics is clearly indicated.}
\label{fig:POD_lisajour_A5d0} 
\end{figure}

Figure \ref{fig:POD_ev_A5d0} shows the energy content of the POD modes for the jet forced at $A=5~$\%. 
The first 8 modes appear in pairs with equal energy, with the first pair capturing more than $90$\% of the total fluctuation kinetic energy. The spatial structure of these first 8 POD modes are shown in figure \ref{fig:POD_modes_A5d0}. All modes are symmetric with respect to the jet axis (by construction). The streamwise component peaks on the axis and the radial component in the shear layer located at approximately $r/D=5$. The mode pairs identified in the energy spectrum correspond to the same spatial structure with a phase lag of $\pi/2$. 
The streamwise wavelengths of the spatial modes indicate that the most energetic modes correspond to the fundamental of the forced oscillation while the less energetic mode pairs represent higher order harmonics. The POD coefficients $a_i$ clearly support this interpretation. The scatter plots shown in figure \ref{fig:POD_lisajour_A5d0} reveal a clear temporal correlation between the first and second, first and fourth, first and sixth, and first and eighth POD coefficient. The circle represents the limit-cycle of the fundamental oscillation, while the single-, double-, and triple-eight-shape of the higher modes reveal that these modes oscillate at a multiple frequency of the fundamental. These scatter plots allow for an nonambiguous assignment of the POD modes to the fundamental flow oscillations and its harmonics. Note again that this information is extracted from the uncorrelated PIV snapshots without any time information.
\par
The clear separation of the POD modes in a fundamental and the higher order harmonics allows for a {\itshape a posteriori} reconstruction of phase-averaged flow quantities, yielding
\begin{subequations}\label{eq:POD_reconstruction}
\begin{align}
 \widetilde{\vect u}_f(\vect x,t)       &= \Re\left\{\hat{\vect u}_f\mathrm{e}^{-i2\pi ft}\right\} = \Re\left\{\overline{\sqrt{a_1^2+a_2^2}}\left(\vect\Phi_1+i\vect\Phi_2\right)\mathrm{e}^{-i2\pi ft}\right\}\\
 \widetilde{\vect u}_{2f}(\vect x,t)    &= \Re\left\{\hat{\vect u}_{2f}\mathrm{e}^{-i4\pi ft}\right\} = \Re\left\{\overline{\sqrt{a_3^2+a_4^2}}\left(\vect\Phi_3+i\vect\Phi_4\right)\mathrm{e}^{-i4\pi ft}\right\}\\
 \widetilde{\vect u}_{3f}(\vect x,t)    &= \Re\left\{\hat{\vect u}_{4f}\mathrm{e}^{-i6\pi ft}\right\} = \Re\left\{\overline{\sqrt{a_5^2+a_6^2}}\left(\vect\Phi_5+i\vect\Phi_6\right)\mathrm{e}^{-i6\pi ft}\right\}...
 \end{align}
\end{subequations}
The amplitude functions $\hat{\vect u}_{f,2f,3f,...}$ are obtained from the corresponding POD mode pairs and are complex numbers that can be expressed as $\hat{\vect u}=|\hat{\vect u}|\exp\left(i\varphi\right)$,
with the amplitude and phase distribution of the fundamental wave given as 
\begin{equation}\label{eq:POD_shapefunction}
|\hat{\vect u}_f| = \overline{\sqrt{a_1^2+a_2^2}}\sqrt{\vect \Phi_1^2+\vect \Phi_2^2} \ \ \ \ \text{and} \ \ \ \   \vect\varphi_f= \arg({\hat{\vect u}_f})=\arctan\left(\frac{\vect \Phi_2}{\vect \Phi_1}\right).
\end{equation}

\section{Weakly nonparallel flow correction}
\label{sec:app1}
\subsection{Governing Equations}
We start with the equations of motion for the perturbation $\vect u'$ linearized around the mean flow $\overline{\vect u}$ as it is given by \eqref{eq:gov_equations_perturbation2}.
These are, for an incompressible, axisymmetric swirling jet, given in cylindrical coordinates as 
\begin{subequations}\label{eq:governing_equations_cylindircal}
\begin{equation}\label{eq:conti_cylindrical}
\frac{1}{r}\frac{\partial r v'}{\partial r} + \frac{1}{r}\frac{\partial w'}{\partial \theta}+\frac{\partial u'}{\partial x}=0
\end{equation}
for continuity, 
\begin{align}\label{eq:axial_cylindrical}
\frac{\partial u'}{\partial t}
+ \left(u'\frac{\partial \overline{u}}{\partial x} + v'\frac{\partial \overline{u}}{\partial r} + \frac{w'}{r}\frac{\partial \overline{u}}{\partial \theta}\right) 
+ \left(\overline{u}\frac{\partial u'}{\partial x} + \overline{v}\frac{\partial u'}{\partial r} + \frac{\overline{w}}{r}\frac{\partial u'}{\partial \theta}\right)
= \nonumber\\
-\frac{\partial p'}{\partial x}
+\frac{1}{\mathrm{Re}}\left(\frac{\partial^2 u'}{\partial x^2}+\frac{\partial^2 u'}{\partial r^2}+\frac{1}{r}\frac{\partial u'}{\partial r}+\frac{1}{r^2}\frac{\partial^2 u'}{\partial \theta^2}\right)
\end{align}
for the axial momentum, 
\begin{align}\label{eq:radial_cylindrical}
\frac{\partial v'}{\partial t}
+ \left(u'\frac{\partial \overline{v}}{\partial x} + v'\frac{\partial \overline{v}}{\partial r} + \frac{w'}{r}\frac{\partial \overline{v}}{\partial \theta} - \frac{w'\overline{w}}{r}\right)  
+ \left(\overline{u}\frac{\partial v'}{\partial x} + \overline{v}\frac{\partial v'}{\partial r} + \frac{\overline{w}}{r}\frac{\partial v'}{\partial \theta}- \frac{w}{r}\right)
= \nonumber\\
 -\frac{\partial p'}{\partial r}
+\frac{1}{\mathrm{Re}}\left(\frac{\partial^2 v'}{\partial x^2}+\frac{\partial^2 v'}{\partial r^2}+\frac{1}{r}\frac{\partial v'}{\partial r}+\frac{1}{r^2}\frac{\partial^2 v'}{\partial \theta^2}-\frac{v'}{r^2}-\frac{2}{r^2}\frac{\partial w'}{\partial \theta}\right)
\end{align}
for the radial momentum, and 
\begin{align}\label{eq:azimithal_cylindrical}
\frac{\partial w'}{\partial t}
+ \left(u'\frac{\partial \overline{w}}{\partial x} + v'\frac{\partial \overline{w}}{\partial r} + \frac{w'}{r}\frac{\partial \overline{w}}{\partial \theta} + \frac{v'\overline{w}}{r}\right) 
+ \left(\overline{u}\frac{\partial w'}{\partial x} + \overline{v}\frac{\partial w'}{\partial r} + \frac{\overline{w}}{r}\frac{\partial w'}{\partial \theta}+ \frac{w'\overline{v}}{r}\right)
= \nonumber\\
 -\frac{1}{r}\frac{\partial p'}{\partial \theta}
+\frac{1}{\mathrm{Re}}\left(\frac{\partial^2 w'}{\partial x^2}+\frac{\partial^2 w'}{\partial r^2}+\frac{1}{r}\frac{\partial w'}{\partial r}+\frac{1}{r^2}\frac{\partial^2 w'}{\partial \theta^2}-\frac{w'}{r^2}+\frac{2}{r^2}\frac{\partial v'}{\partial \theta}\right)
\end{align}
\end{subequations}
for the azimuthal momentum.

\subsection{Multiple-scale ansatz}
For the multiple-scale analysis, we introduce a slow axial scale $X=\epsilon x$, where $\epsilon\ll 1$, and \myred{the} radial component $\overline{v}_1=\overline{v}/\epsilon$. In contrast to \cite{Cooper2002} we do not assume $\epsilon =1/\mathrm{Re}$, so the viscous terms of leading order $O(\epsilon)$ remain. This is consistent with \cite{Reau2002b}. The perturbation is assumed to have the form 
\begin{equation}\label{eq:WKB_perturbation_ansatz1}
[u',v',w',p'](X,r,\theta,t;\epsilon) = [\hat{u},i\hat{v},\hat{w},\hat{p}](X,r;\epsilon)\exp\left(\frac{i}{\epsilon}\int^X\alpha(\xi)\mathrm{d}\xi+im\theta-i\omega t\right). 
\end{equation}
having a slowly varying amplitude and axial wavenumber. 
Partial derivatives of the perturbation with respect to $x$ then become
\begin{equation}
\frac{\partial}{\partial x} = \left(i\alpha(X)+\epsilon\frac{\partial }{\partial X}\right) \quad\quad \text{and} 
\quad\quad 
\frac{\partial^2}{\partial x^2} = \left(\epsilon^2\frac{\partial^2}{\partial X^2}+\epsilon 2i\alpha\frac{\partial}{\partial X}-\alpha^2\myredred{+i\epsilon\frac{\mathrm{d}\alpha}{\mathrm{d}X}}\right).
\end{equation}

By introducing \eqref{eq:WKB_perturbation_ansatz1} into \eqref{eq:governing_equations_cylindircal} and by retaining terms up to $O(\epsilon)$, the linearized equations  governing the disturbance field are 
\begin{subequations}\label{eq:WKB_governing_equations_LHS}
\begin{align}
 \frac{\hat v}{r} + \frac{\partial{\hat v}}{\partial r}+\alpha\hat u+\frac{m}{r}\hat w &= \epsilon f_1 \\
-\frac{i}{\mathrm{Re}}\frac{\partial{^2\hat u}}{\partial r^2}-\frac{i}{\mathrm{Re}~r}\frac{\partial{\hat u}}{\partial{r}}+\left[-i\omega+\frac{im\overline{w}}{r}+i\alpha \overline{u}+\frac{1}{\mathrm{Re}}\left(\frac{m^2}{r^2}+\alpha^2\right)\right]\hat u   \nonumber \\
+ \frac{\partial{\overline{u}}}{\partial{r}}i\hat v + i\alpha\hat p &= \epsilon f_2 \\
-\frac{1}{\mathrm{Re}}\frac{\partial^2{\hat v}}{\partial{r^2}} - \frac{1}{\mathrm{Re}~r}\frac{\partial{\hat v}}{\partial{r}}+\left[\omega-\frac{m\overline{w}}{r}-\alpha \overline{u}+\frac{i}{\mathrm{Re}}
\left(\frac{m^2+1}{r^2}+\alpha^2\right)\right]\hat v \nonumber \\
+\left[\frac{i2m}{\mathrm{Re}~r^2}-\frac{2\overline{w}}{r}\right]\hat w+\frac{\partial{\hat p}}{\partial{r}} &= \epsilon f_3\\
-\frac{1}{\mathrm{Re}}\frac{\partial^2{\hat w}}{\partial{r^2}} - \frac{1}{\mathrm{Re}~r}\frac{\partial{\hat w}}{\partial{r}}+\left[-i\omega+\frac{im\overline{w}}{r}+i\alpha \overline{u}+\frac{1}{\mathrm{Re}}\left(\frac{m^2+1}{r^2}+\alpha^2\right)\right]\hat w   \nonumber \\
\left[i\frac{\partial{\overline{w}}}{\partial{r}}+\frac{2m}{\mathrm{Re}~r^2}+\frac{i\overline{w}}{r}\right]\hat v+\frac{im \hat p}{r}&= \epsilon f_4
\end{align}
\end{subequations}
where the left-hand side (LHS) is equal to the homogeneous local viscous stability problem as formulated by \cite{Khorrami1989}.
  
The functions on the right-hand side  (RHS) are
\begin{subequations}\label{eq:WKB_governing_equations_RHS}
\begin{align}
f_1&=  i \frac{\partial{\hat u}}{\partial{X}}\\
f_2&= -\overline{u}\frac{\partial{\hat u}}{\partial X}-\overline{v}_1\frac{\partial{\hat u}}{\partial{r}}-\frac{\partial{\overline{u}}}{\partial X}\hat u+\frac{i2\alpha}{\mathrm{Re}}\frac{\partial \hat u}{\partial X}\myredred{+\frac{i}{\mathrm{Re}}\frac{\mathrm{d}\alpha}{\mathrm{d}X}\hat u}-\frac{\partial \hat p}{\partial X}\\
f_3&= -i\overline{u}\frac{\partial{\hat v}}{\partial X}-i\overline{v}_1\frac{\partial{\hat v}}{\partial{r}}-i\frac{\partial{\overline{v}_1}}{\partial r}\hat v-\frac{2\alpha}{\mathrm{Re}}\frac{\partial{\hat v}}{\partial X}\myredred{-\frac{1}{\mathrm{Re}}\frac{\mathrm{d}\alpha}{\mathrm{d}X}\hat v}\\
f_4&= -\overline{u}\frac{\partial{\hat w}}{\partial X}-\overline{v}_1\frac{\partial{\hat w}}{\partial{r}}-\frac{\partial{\overline{w}}}{\partial X}\hat u-\frac{\overline{v}_1}{r}\hat w + \frac{i2\alpha}{\mathrm{Re}}\frac{\partial{\hat w}}{\partial X}\myredred{+\frac{i}{\mathrm{Re}}\frac{\mathrm{d}\alpha}{\mathrm{d}X}\hat w}.
\end{align}
\end{subequations}
The set of equations (\ref{eq:WKB_governing_equations_LHS}-\ref{eq:WKB_governing_equations_RHS}) can also be written as
\begin{equation}\label{eq:WKB_Lph}
\mathcal{L}[\hat u, \hat v,\hat w, \hat p] = \epsilon\mathcal{H}[\hat u, \hat v,\hat w, \hat p].
\end{equation}
with the linear operators $\mathcal{L}$ and $\mathcal{H}$ given as
\begin{subequations}\label{eq:WKB_LH}
\begin{equation}
\mathcal{L}=
\begin{pmatrix} 
\alpha& \frac{1}{r}+\frac{\partial}{\partial r} & \frac{m}{r} & 0 \\[1em]
\frac{1}{\mathrm{Re}}(D-\frac{1}{r^2})-iE& i\frac{\partial{\overline{u}}}{\partial{r}} & 0& i\alpha\\[1em]
0& \frac{i}{\mathrm{Re}}D+E& \frac{i2m}{\mathrm{Re}~r^2}-\frac{2\overline{w}}{r} & \frac{\partial}{\partial{r}} \\[1em] 
0& i\frac{\partial{\overline{w}}}{\partial{r}}-\frac{2m}{\mathrm{Re}~r^2}+\frac{i\overline{w}}{r}& \frac{1}{\mathrm{Re}}D-iE & \frac{im}{r}\\[1em]
\end{pmatrix}
\end{equation}
\text{and}\\
\begin{equation}
\mathcal{H}=
\begin{pmatrix} 
i\frac{\partial}{\partial X}  & 0 & 0 & 0\\[1em]
F-\frac{\partial \overline{u}}{\partial X} +\frac{i2\alpha}{\mathrm{Re}}\frac{\partial}{\partial{X}}\myredred{+\frac{i}{\mathrm{Re}}\frac{\mathrm{d}\alpha}{\mathrm{d}X}}&0 & 0 & -\frac{\partial}{\partial X} \\[1em]
0 & iF-i\frac{\partial \overline{v}_1}{\partial r} -\frac{2\alpha}{\mathrm{Re}}\frac{\partial}{\partial{X}}\myredred{-\frac{1}{\mathrm{Re}}\frac{\mathrm{d}\alpha}{\mathrm{d}X}}& 0 & 0 \\[1em]
-\frac{\partial \overline{w}}{\partial X} & 0 & F-\frac{\overline{v}_1}{r}+\frac{i2\alpha}{\mathrm{Re}}\frac{\partial}{\partial{X}}\myredred{+\frac{i}{\mathrm{Re}}\frac{\mathrm{d}\alpha}{\mathrm{d}X}}&0  \\[1em]
\end{pmatrix},
\end{equation}
where 
\begin{align}
E&=\omega-\frac{m}{r}\overline{w}-\alpha \overline{u}\\
D&=-\frac{1}{r}\frac{\partial}{\partial r}\left(r\frac{\partial}{\partial r}\right) + \frac{m^2+1}{r^2}+\alpha^2\\
F&=-\overline{u}\frac{\partial}{\partial X}-\overline{v}_1\frac{\partial}{\partial r}.
\end{align}
\end{subequations}
\subsection{Expansion scheme}
The complex amplitude are now expanded in powers of $\epsilon$ such that 
\begin{equation}\label{eq:WKB_series}
 [\hat u, \hat v,\hat w, \hat p](X,r;\epsilon) = \sum_{n=0}^{\infty} \epsilon^n [\hat u_n, \hat v_n,\hat w_n, \hat p_n](X,r).   
\end{equation}
This expansion series is substituted into \eqref{eq:WKB_LH}.
We are interested in first order ($n=1$), yielding
\begin{equation}\label{eq:WKB_order1}
\mathcal{L} 
\begin{pmatrix} 
\hat u_0 \\
\hat v_0\\
\hat w_0 \\
\hat p_0
\end{pmatrix}
+
 \epsilon\mathcal{L} 
\begin{pmatrix} 
 \hat u_1 \\
\hat v_1 \\
\hat w_1 \\
\hat p_1 
\end{pmatrix}
=
\epsilon\mathcal{H} 
\begin{pmatrix} 
\hat u_0\\
\hat v_0\\
\hat w_0\\
\hat p_0 
\end{pmatrix}
+
\epsilon^2\mathcal{H} 
\begin{pmatrix} 
\hat u_1 \\
\hat v_1 \\
\hat w_1 \\
\hat p_1 
\end{pmatrix}.
\end{equation}
At leading order, the solution of $\mathcal{L}[\hat u_0, \hat v_0, \hat w_0, \hat p_0]=0$ represents the stability modes of the parallel flow. 
Since $\mathcal{L}$ does not contain any $X$-derivatives, the solutions can be multiplied by an arbitrary function in $X$. Hence, if $[\check u_0, \check v_0, \check w_0, \check p_0]$ is the solution of the homogeneous problem, a general solution is 
\begin{equation}\label{eq:WKB_Nx}
\begin{pmatrix} 
\hat u_0\\
\hat v_0\\
\hat w_0\\
\hat p_0 
\end{pmatrix}
= N(X)
\begin{pmatrix} 
\check u_0\\
\check v_0\\
\check w_0\\
\check p_0 
\end{pmatrix}.
\end{equation}

The scaling $N(X)$ depends on how the homogeneous solutions are normalized and it connects the solutions in streamwise direction.
It is derived by the fact that the first-order equations are solvable and that their inhomogeneous terms, which represent the dependence on $X$, are functions of the leading order (homogeneous) solutions only. 
Hence, the solvability  of the first-order equation $\mathcal{L}[\hat u_1, \hat v_1, \hat w_1, \hat p]=\mathcal{H}[\hat u_0, \hat v_0, \hat w_0, \hat p_0]$ puts a constrain on the spatial derivatives of the zero-order solutions.
\par
The solvability condition is obtained by taking the inner product of the first-order equations with the adjoint solution of the zero-order (homogeneous) problem. 
The inner product is defined as
\begin{equation}
 \left\langle 
\begin{pmatrix} 
\hat u_1\\
\hat v_1\\
\hat w_1\\
\hat p_1 
\end{pmatrix},
\begin{pmatrix} 
\hat u_2\\
\hat v_2\\
\hat w_2\\
\hat p_2 
\end{pmatrix}
\right\rangle
= \int_0^{\infty}(\underaccent{\bar}{\hat u}_1\hat u_2+\underaccent{\bar}{\hat v}_1\hat v_2  +\underaccent{\bar}{\hat w}_1\hat w_2  +\underaccent{\bar}{\hat p}_1\hat p_2  )r{\mathrm d}r,
\end{equation}
where the bar under the symbol denotes the complex conjugate.
The adjoint solution is obtained from the adjoint homogeneous problem 
\begin{equation}\label{eq:WKB_adjoint}
 \mathcal{L}^*
\begin{pmatrix} 
\check u^*_0\\
\check v^*_0\\
\check w^*_0\\
\check p^*_0 
\end{pmatrix}
=0.
\end{equation}
The solvability condition to determine $N(X)$ is obtained from the relation
\begin{equation}
\left\langle 
\begin{pmatrix} 
\check u_0^*\\
\check v_0^*\\
\check w_0^*\\
\check p_0^* 
\end{pmatrix} 
,
\mathcal{L}
\begin{pmatrix} 
\hat u_1^*\\
\hat v_1^*\\
\hat w_1^*\\
\hat p_1^* 
\end{pmatrix} 
\right\rangle
=
\left\langle 
\mathcal{L}^*
\begin{pmatrix} 
\check u_0^*\\
\check v_0^*\\
\check w_0^*\\
\check p_0^* 
\end{pmatrix} 
,
\begin{pmatrix} 
\hat u_1\\
\hat v_1\\
\hat w_1\\
\hat p_1 
\end{pmatrix} 
\right\rangle
\end{equation}
which simplifies to
\begin{equation}\label{eq:WKB_scalaproduct}
\left\langle 
\begin{pmatrix} 
\check u_0^*\\
\check v_0^*\\
\check w_0^*\\
\check p_0^* 
\end{pmatrix} 
,
\mathcal{H}
\begin{pmatrix} 
\hat u_0\\
\hat v_0\\
\hat w_0\\
\hat p_0 
\end{pmatrix} 
\right\rangle
=0.
\end{equation}
The terms $\mathcal{H}[\hat u_0, \hat v_0,\hat w_0,\hat p_0]$ consist of terms involving disturbances and axial derivatives of disturbances, and, with relation \eqref{eq:WKB_Nx}, they can be rearranged to 
\begin{equation}
\mathcal{H}
\begin{pmatrix} 
\hat u_0\\
\hat v_0\\
\hat w_0\\
\hat p_0 
\end{pmatrix} 
= 
\frac{\mathrm{d}N}{\mathrm{d}X}
\mathcal{H}_1
\begin{pmatrix} 
\check u_0\\
\check v_0\\
\check w_0\\
\check p_0 
\end{pmatrix}
+
N(X)
\mathcal{H}_2
\begin{pmatrix} 
\check u_0\\
\check v_0\\
\check w_0\\
\check p_0 
\end{pmatrix}
\end{equation}
with the operators $\mathcal{H}_1$ and $\mathcal{H}_2$ to be
\begin{equation}
\mathcal{H}_1
=
\begin{pmatrix} 
i& 0 &0& 0\\[1em]
-\overline{u}+\frac{2i\alpha}{\mathrm{Re}}& 0 &0& -1\\[1em]
0& -i\overline{u}-\frac{2\alpha}{\mathrm{Re}} &0& 0\\[1em]
0& 0 &-\overline{u}+\frac{2i\alpha}{\mathrm{Re}}& 0
\end{pmatrix}
\end{equation}

and

\begin{equation}
\mathcal{H}_2
=
\begin{pmatrix} 
i\frac{\partial}{\partial X}  & 0 & 0 & 0\\[1em]
F-\frac{\partial \overline{u}}{\partial X}+\frac{i2\alpha}{\mathrm{Re}}\frac{\partial}{\partial{X}}\myredred{+\frac{i}{\mathrm{Re}}\frac{\mathrm{d}\alpha}{\mathrm{d}X}} &0 & 0 & -\frac{\partial}{\partial X} \\[1em]
0 & iF-i\frac{\partial \overline{v}_1}{\partial r}-\frac{2\alpha}{\mathrm{Re}}\frac{\partial}{\partial{X}}\myredred{-\frac{1}{\mathrm{Re}}\frac{\mathrm{d}\alpha}{\mathrm{d}X}}& 0 & 0 \\[1em]
-\frac{\partial \overline{w}}{\partial X} & 0 & F-\frac{\overline{v}_1}{r}+\frac{i2\alpha}{\mathrm{Re}}\frac{\partial}{\partial{X}}\myredred{+\frac{i}{\mathrm{Re}}\frac{\mathrm{d}\alpha}{\mathrm{d}X}}&0
  \end{pmatrix}.
\end{equation}

The scalar product \eqref{eq:WKB_scalaproduct} becomes then
\begin{equation}
\frac{\mathrm{d}N}{\mathrm{d}X}
\underbrace{
\left\langle
\begin{pmatrix} 
\check u_0^*\\
\check v_0^*\\
\check w_0^*\\
\check p_0^* 
\end{pmatrix}
,
\mathcal{H}_1
\begin{pmatrix} 
\check u_0\\
\check v_0\\
\check w_0\\
\check p_0 
\end{pmatrix}
\right\rangle
}_{G}
+
N(X)
\underbrace{
\left\langle
\begin{pmatrix} 
\check u_0^*\\
\check v_0^*\\
\check w_0^*\\
\check p_0^* 
\end{pmatrix} 
,
\mathcal{H}_2
\begin{pmatrix} 
\check u_0\\
\check v_0\\
\check w_0\\
\check p_0 
\end{pmatrix}
\right\rangle
}_{K}
=0
\end{equation}
with 
\begin{equation}\label{eq:G}
 G = \int_0^{\infty}\left(
i\underaccent{\bar}{\check u}_0^*\check u_0
+ \left[-\overline{u}+\frac{2i\alpha}{\mathrm{Re}}\right]\left(\underaccent{\bar}{\check v}_0^*\check u_0 
+ \underaccent{\bar}{\check w}_0^*i\check v_0
+ \underaccent{\bar}{\check p}_0^*\check w_0\right)-\underaccent{\bar}{\check v}_0^*\check p_0 
\right)r\mathrm{d}r
\end{equation}
and
\begin{align}\label{eq:K}
 K = \int_0^{\infty}\left(i\underaccent{\bar}{\check u}_0^*\frac{\partial \check u_0}{\partial{X}}
+\underaccent{\bar}{\check v}_0^*\left[F(\check u_0)-\frac{\partial \overline{u}}{\partial X}\check u_0+\frac{2i\alpha}{\mathrm{Re}}\frac{\partial\check u_0}{\partial X}\myredred{+\frac{i}{\mathrm{Re}}\frac{\mathrm{d}\alpha}{\mathrm{d}X}\check u_0}-\frac{\partial \check p_0}{\partial X}\right]\right. \nonumber \\ \left. 
+\underaccent{\bar}{\check w}_0^*\left[iF(\check v_0)-i\frac{\partial \overline{v}_1}{\partial r}\check v_0-\frac{2\alpha}{\mathrm{Re}}\frac{\partial\check v_0}{\partial X}\myredred{-\frac{1}{\mathrm{Re}}\frac{\mathrm{d}\alpha}{\mathrm{d}X}\check v_0}\right] \right. \nonumber \\ \left. 
+\underaccent{\bar}{\check p}_0^*\left[-\frac{\partial \overline{w}}{\partial X}\check u_0+F(\check w_0)-\frac{\overline{v}_1}{r}\check w_0+\frac{2i\alpha}{\mathrm{Re}}\frac{\partial\check w_0}{\partial X}\myredred{+\frac{i}{\mathrm{Re}}\frac{\mathrm{d}\alpha}{\mathrm{d}X}\check w_0}\right]
\right)r\mathrm{d}r.
\end{align}
This can be written as 
$$
G(X)\frac{\mathrm{d}N(X)}{\mathrm{d}X} = -K(X)N(X)
$$
and the amplitude scaling is then given by
\begin{equation}\label{eq:WKB_Nx_result}
 N(X) = N_0 \exp\left(\int^X-\frac{K(\xi)}{G(\xi)}\mathrm{d}\xi\right)
\end{equation}
where $N_0$ is an arbitrary normalization constant.
By substituting \eqref{eq:WKB_Nx} and \eqref{eq:WKB_Nx_result} in \eqref{eq:WKB_perturbation_ansatz1}, we obtain an expression for the perturbation 
\begin{align*}
 [u,v,w,p]&(X,r,\theta,t;\epsilon) = \\
 &N_0 [\check u_0 ,\check v_0,\check w_0,\check p_0](X,r;\epsilon)\exp\left(\frac{i}{\epsilon}\int^X\alpha(\xi)\mathrm{d}\xi-\int^X\frac{K(\xi)}{G(\xi)}\mathrm{d}\xi+im\theta-i\omega t\right). 
\end{align*}
Using $\overline{v}=\epsilon \overline{v}_1$ and $\epsilon\frac{\partial}{\partial X}=\frac{\partial}{\partial x}$ we can drop the formal expansion parameter $\epsilon$, yielding
\begin{align*}
 [u,v,w,p](x,&r,\theta,t) = \\
 &N_0 [\check u_0 ,\check v_0,\check w_0,\check p_0](x,r)\exp\left(i\int^x\left[\alpha(\xi)+i\frac{K(\xi)}{G(\xi)}\right]\mathrm{d}\xi+im\theta-i\omega t\right). 
\end{align*}

\subsection{Computing the direct and adjoint local eigenvalues}
\todo{kann eventuell im haupttext untergebracht werden}
At leading order, the set of equations (\ref{eq:WKB_governing_equations_LHS}-\ref{eq:WKB_governing_equations_RHS}) can be formulated as an eigenvalue problem  
\begin{equation}\label{eq:EVP}
\mathbf A(\omega)\vect \psi_0=\alpha\mathbf B(\omega)\vect \psi_0,
\end{equation}
with the eigenvalue $\alpha$ and eigenfunction $\vect \psi_0=(\hat{u}_0,\hat{v}_0,\hat{w}_0,\hat{p}_0)^T$, and the matrix $\mathbf A(\omega)$ and $\mathbf B(\omega)$ containing the mean flow profiles. It is solved for the boundary conditions at $r=\infty$ \citep{Khorrami1989} 
\begin{equation}\label{bc_inf}
\hat{u}_0=\hat{v}_0=\hat{w}_0=\hat{p}_0=0
\end{equation}
and in the limit along the centerline $(r=0)$ 
\begin{subequations}\label{eq:bc_01}
\begin{align}
\hat{u}_0=\hat{v}_0=\hat{w}_0=\hat{p}_0=0 \quad &\hbox{if}  \quad &|m|&>1\\
\left.\begin{aligned}\hat{w}_0=\hat{p}_0&=0\\ 
\hat{v}_0+m\hat{u}_0&=0\\
2\mathrm{d}\hat{v}_0/\mathrm{d}r+m\mathrm{d}\hat{u}_0/\mathrm{d}r&=0
\end{aligned}\quad\right\} \quad &\hbox{if} \quad &|m|&=1\\
\left.\begin{aligned}\hat{v}_0(0)=\hat{w}_0(0)=0\,\\ 
\hat{u}_0 \quad \hbox{and}  \quad \hat{p}_0 \quad  \hbox{finite}\, 
\end{aligned}\quad\right\} \quad &\hbox{if} \quad &m\:&=0.
\end{align}
\end{subequations}
The eigenvalue problem was discretized using a Chebyshev spectral collocation
method and solved directly in MATLAB\texttrademark. A detailed description of the numerical approach
is given in \cite{Oberleithner2011a} and \cite{Oberleithner2013c}.
The adjoint eigenvector $\underaccent{\bar}{\hat{\vect v}}$ is derived by solving the adjoint eigenvalue problem 
\begin{equation}\label{eq:EVPad}
\mathbf A^H(\omega) \underaccent{\bar}{\hat{ \vect v}}=\underaccent{\bar}{\alpha}\mathbf B^H(\omega)\underaccent{\bar}{\hat{\vect v}},
\end{equation}
with $\underaccent{\bar}{\alpha}=\alpha^*$. The superscript $H$ denotes the conjugate transpose and the asterisk denotes the conjugate complex.

\bibliographystyle{jfm2}


\begin{thebibliography}{41}
\expandafter\ifx\csname natexlab\endcsname\relax\def\natexlab#1{#1}\fi

\bibitem[{Barkley}(2006)]{Barkley2006}
{\sc {Barkley}, D.} 2006 {Linear analysis of the cylinder wake mean flow}. {\em
  Europhys. Lett.\/} {\bf 75}, 750--756.

\bibitem[{Berkooz} {\em et~al.\/}(1993){Berkooz}, {Holmes} \&
  {Lumley}]{Berkooz1993}
{\sc {Berkooz}, G., {Holmes}, P. \& {Lumley}, J.~L.} 1993 {The proper
  orthogonal decomposition in the analysis of turbulent flows}. {\em Annu. Rev.
  Fluid Mech.\/} {\bf 25}, 539--575.

\bibitem[Brown \& Roshko(1974)]{Brown1974}
{\sc Brown, G.~L. \& Roshko, A.} 1974 {On density effects and large structure
  in turbulent mixing layers}. {\em J. Fluid Mech.\/} {\bf 64}, 775--816.

\bibitem[{Cater} \& {Soria}(2002)]{Cater2002}
{\sc {Cater}, J.~E. \& {Soria}, J.} 2002 {The evolution of round
  zero-net-mass-flux jets}. {\em J. Fluid Mech.\/} {\bf 472}, 167--200.

\bibitem[Cohen \& Wygnanski(1987)]{Cohen1987a}
{\sc Cohen, J. \& Wygnanski, I.} 1987 The evolution of instabilities in the
  axisymmetric jet. part 1. the linear growth of disturbances near the nozzle.
  {\em J. Fluid Mech.\/} {\bf 176}, 191--219.

\bibitem[{Cooper} \& {Peake}(2002)]{Cooper2002}
{\sc {Cooper}, A.~J. \& {Peake}, N.} 2002 {The stability of a slowly diverging
  swirling jet}. {\em J. Fluid Mech.\/} {\bf 473}, 389--411.

\bibitem[{Crighton} \& {Gaster}(1976)]{Crighton1976}
{\sc {Crighton}, D.~G. \& {Gaster}, M.} 1976 {Stability of slowly diverging jet
  flow}. {\em J. Fluid Mech.\/} {\bf 77}, 397--413.

\bibitem[{Crow} \& {Champagne}(1971)]{Crow1971}
{\sc {Crow}, S.~C. \& {Champagne}, F.~H.} 1971 {Orderly structure in jet
  turbulence}. {\em J. Fluid Mech.\/} {\bf 48}, 547--591.

\bibitem[Edgington-Mitchell {\em et~al.\/}(2014)Edgington-Mitchell,
  Oberleithner, Honnery \& Soria]{Mitchell2014}
{\sc Edgington-Mitchell, D., Oberleithner, K., Honnery, D.~R. \& Soria, J.}
  2014 {Coherent structure and sound production in the helical mode of a
  screeching axisymmetric jet}. {\em J. Fluid Mech.\/} {\bf 748}, 822--847.

\bibitem[von Ellenrieder {\em et~al.\/}(2001)von Ellenrieder, Kostas \&
  Soria]{vonEllenrieder2001}
{\sc von Ellenrieder, K., Kostas, J. \& Soria, J.} 2001 Measurements of a
  wall-bounded turbulent, separated flow using {HPIV}. {\em Journal of
  Turbulence\/} {\bf 2}, 1--15.

\bibitem[{Gaster} {\em et~al.\/}(1985){Gaster}, {Kit} \&
  {Wygnanski}]{Gaster1985}
{\sc {Gaster}, M., {Kit}, E. \& {Wygnanski}, I.} 1985 {Large-scale structures
  in a forced turbulent mixing layer}. {\em J. Fluid Mech.\/} {\bf 150},
  23--39.

\bibitem[Greenblatt \& Wygnanski(2000)]{Greenblatt2000}
{\sc Greenblatt, D. \& Wygnanski, I.~J.} 2000 The control of flow separation by
  periodic excitation. {\em Progress in Aerospace Sciences\/} {\bf 36}~(7), 487
  -- 545.

\bibitem[Gudmundsson \& Colonius(2011)]{Gudmundsson2011}
{\sc Gudmundsson, K. \& Colonius, T.} 2011 Instability wave models for the
  near-field fluctuations of turbulent jets. {\em J. Fluid Mech.\/} {\bf 689},
  97--128.

\bibitem[Huang {\em et~al.\/}(1993)Huang, Fiedler \& Wang]{Huang1993}
{\sc Huang, H.~T., Fiedler, H.~E. \& Wang, J.~J.} 1993 {Limitation and
  improvement of PIV}. {\em Exp. Fluids\/} {\bf 15-15}~(4-5), 263--273.

\bibitem[{Huerre} \& {Monkewitz}(1990)]{Huerre1990}
{\sc {Huerre}, P. \& {Monkewitz}, P.~A.} 1990 {Local and global instabilities
  in spatially developing flows}. {\em Annu. Rev. Fluid Mech.\/} {\bf 22},
  473--537.

\bibitem[{Juniper} {\em et~al.\/}(2011){Juniper}, {Tammisola} \&
  {Lundell}]{Juniper2011a}
{\sc {Juniper}, M.~P., {Tammisola}, O. \& {Lundell}, F.} 2011 {The local and
  global stability of confined planar wakes at intermediate Reynolds number}.
  {\em J. Fluid Mech.\/} {\bf 686}, 218--238.

\bibitem[Khorrami {\em et~al.\/}(1989)Khorrami, Malik \& Ash]{Khorrami1989}
{\sc Khorrami, M.~R., Malik, M.~R. \& Ash, R.~L.} 1989 Application of spectral
  collocation techniques to the stability of swirling flows. {\em J. Comput.
  Phys.\/} {\bf 81}~(1), 206--229.

\bibitem[Lifshitz {\em et~al.\/}(2008)Lifshitz, Degani \& Tumin]{Lifshitz2008}
{\sc Lifshitz, Y., Degani, D. \& Tumin, A.} 2008 On the interaction of
  turbulent shear layers with harmonic perturbations. {\em Flow, Turbulence and
  Combustion\/} {\bf 80}~(1), 61--80.

\bibitem[{Marasli} {\em et~al.\/}(1991){Marasli}, {Champagne} \&
  {Wygnanski}]{Marasli1991}
{\sc {Marasli}, B., {Champagne}, F.~H. \& {Wygnanski}, I.~J.} 1991 {On linear
  evolution of unstable disturbances in a plane turbulent wake}. {\em Phys.
  Fluids\/} {\bf 3}, 665--674.

\bibitem[Meliga {\em et~al.\/}(2012)Meliga, Pujals \& \'{E}ric
  Serre]{Meliga2012a}
{\sc Meliga, P., Pujals, G. \& \'{E}ric Serre} 2012 Sensitivity of 2-d
  turbulent flow past a d-shaped cylinder using global stability. {\em Phys.
  Fluids\/} {\bf 24}~(6), 061701.

\bibitem[Noack {\em et~al.\/}(2003)Noack, Afanasiev, Morzy\'nski, Tadmor \&
  Thiele]{Noack2003}
{\sc Noack, B.~R., Afanasiev, K., Morzy\'nski, M., Tadmor, G. \& Thiele, F.}
  2003 A hierarchy of low-dimensional models for the transient and
  post-transient cylinder wake. {\em J. Fluid Mech.\/} {\bf 497}, 335--363.

\bibitem[{Oberleithner} {\em et~al.\/}(2012){Oberleithner}, {Paschereit},
  {Seele} \& {Wygnanski}]{Oberleithner2012b}
{\sc {Oberleithner}, K., {Paschereit}, C.~O., {Seele}, R. \& {Wygnanski}, I.}
  2012 {Formation of Turbulent Vortex Breakdown: Intermittency, Criticality,
  and Global Instability}. {\em AIAA Journal\/} {\bf 50}, 1437--1452.

\bibitem[Oberleithner {\em et~al.\/}(2014)Oberleithner, Paschereit \&
  Wygnanski]{Oberleithner2013c}
{\sc Oberleithner, K., Paschereit, C.~O. \& Wygnanski, I.} 2014 {On the impact
  of swirl on the growth of coherent structures}. {\em J. Fluid Mech.\/} {\bf
  741}, 156--199.

\bibitem[Oberleithner {\em et~al.\/}(2011)Oberleithner, Sieber, Nayeri,
  Paschereit, Petz, Hege, Noack \& Wygnanski]{Oberleithner2011a}
{\sc Oberleithner, K., Sieber, M., Nayeri, C.~N., Paschereit, C.~O., Petz, C.,
  Hege, H.-C., Noack, B.~R. \& Wygnanski, I.} 2011 Three-dimensional coherent
  structures in a swirling jet undergoing vortex breakdown: stability analysis
  and empirical mode construction. {\em J. Fluid Mech.\/} {\bf 679}, 383--414.

\bibitem[{O'Neill} {\em et~al.\/}(2004){O'Neill}, {Soria} \&
  {Honnery}]{ONeill2004}
{\sc {O'Neill}, P., {Soria}, J. \& {Honnery}, D.} 2004 {The stability of low
  Reynolds number round jets}. {\em Exp. Fluids\/} {\bf 36}, 473--483.

\bibitem[Orszag \& Crow(1970)]{Orszag1970}
{\sc Orszag, S.~A., S. \& Crow, S.~C., S.} 1970 Instability of a vortex sheet
  leaving a semi-infinite plate(vortex sheet instability leaving semiinfinite
  flat plate, considering boundary effects). {\em Stud. Appl. Math.\/} {\bf
  49}, 167--181.

\bibitem[{Pier}(2002)]{Pier2002}
{\sc {Pier}, B.} 2002 {On the frequency selection of finite-amplitude vortex
  shedding in the cylinder wake}. {\em J. Fluid Mech.\/} {\bf 458}, 407--417.

\bibitem[{Reau} \& {Tumin}(2002)]{Reau2002b}
{\sc {Reau}, N. \& {Tumin}, A.} 2002 {Harmonic Perturbations in Turbulent
  Wakes}. {\em AIAA Journal\/} {\bf 40}, 526--530.

\bibitem[{Reynolds} \& {Hussain}(1972)]{Reynolds1972}
{\sc {Reynolds}, W.~C. \& {Hussain}, A.~K.~M.~F.} 1972 {The mechanics of an
  organized wave in turbulent shear flow. Part 3. Theoretical models and
  comparisons with experiments}. {\em J. Fluid Mech.\/} {\bf 54}, 263--288.

\bibitem[Rienstra(1983)]{Rienstra1983}
{\sc Rienstra, S.} 1983 A small strouhal number analysis for acoustic wave-jet
  flow-pipe interaction. {\em Journal of Sound and Vibration\/} {\bf 86}~(4),
  539--556.

\bibitem[{Sipp} \& {Lebedev}(2007)]{Sipp2007a}
{\sc {Sipp}, D. \& {Lebedev}, A.} 2007 {Global stability of base and mean
  flows: a general approach and its applications to cylinder and open cavity
  flows}. {\em J. Fluid Mech.\/} {\bf 593}, 333--358.

\bibitem[{Sipp} {\em et~al.\/}(2010){Sipp}, {Marquet}, {Meliga} \&
  {Barbagallo}]{Sipp2010a}
{\sc {Sipp}, D., {Marquet}, O., {Meliga}, P. \& {Barbagallo}, A.} 2010
  {Dynamics and Control of Global Instabilities in Open-Flows: A Linearized
  Approach}. {\em Applied Mechanics Reviews\/} {\bf 63}~(3), 030801.

\bibitem[Sirovich(1987)]{Sirovich1987}
{\sc Sirovich, L.} 1987 Turbulence and the dynamics of coherent structures.
  part i: Coherent structures. {\em Quarterly of Applied Mathematics\/} {\bf
  XLV}, 561--571.

\bibitem[Soria(1994)]{Soria1994c}
{\sc Soria, J.} 1994 Digital cross-correlation particle image velocimetry
  measurements in the near wake of a circular cylinder. In {\em International
  Colloquium on Jets, Wakes and Shear Layers\/}. Melbourne, Australia.

\bibitem[Soria(1996{\natexlab{{\em a\/}}})]{Soria1996c}
{\sc Soria, J.} 1996{\natexlab{{\em a\/}}} An adaptive cross-correlation
  digital {PIV} technique for unsteady flow investigations. In {\em Proc. 1st
  Australian Conference on Laser Diagnostics in Fluid Mechanics and
  Combustion\/} (ed. A.~Masri \& D.~Honnery), pp. 29 -- 48. University of
  Sydney, University of Sydney, Sydney, NSW, Australia.

\bibitem[Soria(1996{\natexlab{{\em b\/}}})]{Soria1996b}
{\sc Soria, J.} 1996{\natexlab{{\em b\/}}} An investigation of the near wake of
  a circular cylinder using a video-based digital cross-correlation particle
  image velocimetry technique. {\em Exp. Therm Fluid Sci.\/} {\bf 12}, 221 --
  233.

\bibitem[Soria(1998)]{Soria1998b}
{\sc Soria, J.} 1998 Multigrid approach to cross-correlation digital {PIV} and
  {HPIV} analysis. In {\em Proceedings of 13th Australasian Fluid Mechanics
  Conference\/}. Monash University, Melbourne, Australia.

\bibitem[Soria {\em et~al.\/}(1999)Soria, Cater \& Kostas]{Soria1999}
{\sc Soria, J., Cater, J. \& Kostas, J.} 1999 High resolution multigrid
  cross-correlation digital {PIV} measurements of a turbulent starting jet
  using half frame image shift film recording. {\em Optics and Laser
  Technology\/} {\bf 31}, 3--12.

\bibitem[St{\"o}hr {\em et~al.\/}(2011)St{\"o}hr, Sadanandan \&
  Meier]{Stoehr2011a}
{\sc St{\"o}hr, M., Sadanandan, R. \& Meier, W.} 2011 Phase-resolved
  characterization of vortex�flame interaction in a turbulent swirl flame.
  {\em Exp. Fluids\/} {\bf 51}, 1153--1167. 10.1007/s00348-011-1134-y.

\bibitem[{Strange} \& {Crighton}(1983)]{Strange1983}
{\sc {Strange}, P.~J.~R. \& {Crighton}, D.~G.} 1983 {Spinning modes on
  axisymmetric jets. I}. {\em J. Fluid Mech.\/} {\bf 134}, 231--245.

\bibitem[{Weisbrot} \& {Wygnanski}(1988)]{Weisbrot1988}
{\sc {Weisbrot}, I. \& {Wygnanski}, I.} 1988 {On coherent structures in a
  highly excited mixing layer}. {\em J. Fluid Mech.\/} {\bf 195}, 137.

\end{thebibliography}

\end{document}